\documentclass[11pt]{article}
\newcommand{\be}{\begin{equation}}
\newcommand{\ee}{\end{equation}}
\newcommand{\bea}{\begin{eqnarray}}
\newcommand{\eea}{\end{eqnarray}}
\newcommand{\sn}{{\rm sn}}

\newcommand{\dn}{{\rm dn}}
\newcommand{\cn}{{\rm cn}}
\newcommand{\sech}{{\rm sech}}

\begin{document}
\vspace{0.5in}

\begin{center}
{\LARGE{\bf New Solutions of Coupled Nonlocal NLS and Coupled 
Nonlocal mKdV Equations}}
\end{center}

\begin{center}
{\LARGE{\bf Avinash Khare}} \\
{Physics Department, Savitribai Phule Pune University \\
 Pune 411007, India}
\end{center}

\begin{center}
{\LARGE{\bf Avadh Saxena}} \\ 
{Theoretical Division and Center for Nonlinear Studies, 
Los Alamos National Laboratory, Los Alamos, New Mexico 87545, USA}
\end{center}

\begin{abstract}
We provide several novel solutions of the coupled Ablowitz-Musslimani (AM) 
version of the nonlocal nonlinear Schr\"odinger (NLS) equation and the coupled 
nonlocal modified Korteweg-de Vries (mKdV) equations. In each case we 
compare and contrast the corresponding solutions of the relevant coupled 
local equations. Further, we provide new solutions of the coupled local NLS
and the coupled local mKdV equations which are not the solutions of the 
corresponding nonlocal equations.
\end{abstract}


\section{Introduction}

In recent years the nonlocal nonlinear equations have received considerable
attention in the literature. It primarily started with the papers of 
Ablowitz and Musslimani (AM) \cite{am1, am2} about the nonlocal NLS equation 
and its integrability \cite{gerd}. Subsequently, several other nonlocal
equations have been proposed \cite{mkdv,gur,cen,hirota,yang}. It is hoped 
that some of these nonlocal equations may have applications in areas of physics  
such as optics, photonics, etc. \cite{optics}. Besides one hopes to understand the 
nature of the relevant nonlocality. It is clearly possible that in some cases 
coupled nonlocal equations could be relevant in a number of physical settings. 

It is with this motivation that recently we have studied coupled AM type 
nonlocal NLS as well as nonlocal coupled mKdV equations \cite{ks14,ks15,ks22}
and obtained a large number of solutions of these coupled equations. We thought 
that perhaps one has essentially covered most of the allowed solutions. 
Of course, we were aware that nonlinear equations being rich one can never be
sure about it. Recently, we were able to obtain new solutions of the 
uncoupled symmetric $\phi^4$ equation \cite{kbs} as well as well as the coupled
$\phi^4$ equations \cite{kbs1} and that has inspired us to look for similar 
solutions of the coupled nonlocal equations. The purpose of this paper is to report 
on several new solutions of the coupled AM variant of the nonlocal NLS as well as 
the coupled nonlocal mKdV equations \cite{mkdv,gur}. In each case we also 
enquire if the corresponding local equations also admit such a solution, and if yes,  
under what conditions. 

The plan of the paper is the following. In Sec. II we present new solutions 
of the coupled nonlocal AM variant of the NLS equation. In each case we
compare and contrast with the solutions of the (local) coupled NLS equations
(in case they are admitted). Particular mention may be made of a truncated 
coupled nonlocal NLS equation which admits novel nonreciprocal solutions 
where one field is in terms of the Lam\'e polynomial of order two while the other 
field is in terms of the Lam\'e polynomial of order one. It may be noted that the 
full coupled nonlocal or local NLS equations do not admit similar solutions.  
In Appendix A we present those solutions 
of the local coupled NLS equations which are not the solutions of the 
nonlocal coupled NLS equations. In Sec. III we present new 
solutions of the nonlocal coupled mKdV equations and compare and contrast them 
with the corresponding solutions of the local coupled mKdV equations. 
Particular mention may be made of the plane wave solutions as well as the 
soliton solutions multiplied by the plane wave factor of the coupled 
nonlocal mKdV equations. In contrast, the corresponding local coupled mKdV
equations do not admit either the plane wave or the soliton solutions 
multiplied by the plane wave factor.
In Appendix B we present those solutions of the local coupled 
mKdV equations which are not the solutions of the nonlocal coupled mKdV. 

\section{Novel Solutions of a Coupled nonlocal Ablowitz-Musslimani NLS Model}

Let us consider the following coupled nonlocal Ablowitz-Musslimani NLS
model \cite{ks15,ks22}
\be\label{1}
iu_t(x,t) +u_{xx}(x,t) +[g_{11} u(x,t) u^{*}(-x,t) 
+g_{12}v(x,t) v^{*}(-x,t)]u(x,t) = 0\,,
\ee
\be\label{2} 
iv_t(x,t) +v_{xx}(x,t) +[g_{21} u(x,t) u^{*}(-x,t) 
+g_{22}v(x,t) v^{*}(-x,t)]v(x,t) = 0\,. 
\ee
In order to find exact solutions to the coupled Eqs. (\ref{1}) and 
(\ref{2}), we start with the ansatz
\be\label{3}
u(x,t) = e^{i\omega_1(t+t_0)} u(x)\,,~~v(x,t) = e^{i\omega_2(t+t_1)}\,,
\ee
where $t_0, t_1$ are arbitrary real constants. In that case Eqs. (\ref{1})
and (\ref{2}) take the form
\be\label{4}
u_{xx}(x) = \omega_1 u(x) -[g_{11} u(x) u^{*}(-x) 
+g_{12}v(x) v^{*}(-x)]u(x)\,,
\ee
\be\label{5} 
v_{xx}(x) = \omega_2 v(x) -[g_{21} u(x) u^{*}(-x) 
+g_{22}v(x,t) v^{*}(-x)]v(x)\,. 
\ee
For simplicity, now onwards we will put $t_0, t_1 = 0$. However, it is
understood that such a shift in time $t$ is always allowed. However, in 
view of the nonlocality, similar shift in $x$ is not allowed. This is 
unlike the local NLS case where such a shift is allowed.

We now discuss new solutions of these coupled equations and point out
whether the corresponding coupled local NLS equations
\be\label{6}
iu_t(x,t) +u_{xx}(x,t) +[g_{11} u^2(x,t) 
+g_{12}v^2(x,t)]u(x,t) = 0\,,
\ee
\be\label{7} 
iv_t(x,t) +v_{xx}(x,t) +[g_{21} u^2(x,t) 
+g_{22}v^2(x,t)]v(x,t) = 0\,,
\ee
also admit these solutions and if yes under what conditions. In Appendix A, 
we discuss those solutions which are admitted by the local coupled
Eqs. (\ref{6}) and (\ref{7}) but not by the coupled nonlocal Eqs. (\ref{1})
and (\ref{2}).

{\bf Solution I}

It is not difficult to check that
\be\label{8}
u(x,t) = e^{i\omega_1 t} \frac{A\sqrt{m} \sn(\beta x,m)}{B+\dn(\beta x, m)}\,,~~
v(x,t) = e^{i\omega_2 t} \frac{D\sqrt{m} \cn(\beta x,m)}{B+\dn(\beta x, m)}\,,
\ee
with $B > 0$, is an exact solution of the coupled Eqs. (\ref{1}) 
and (\ref{2}) provided
\bea\label{9}
&&\omega_1 = \omega_2 = -\frac{(2-m)\beta^2}{2}\,,~~g_{12} D^2 = 
3 g_{22} D^2 = \frac{3(B^2-1)\beta^2}{2}\,, \nonumber \\
&&3 g_{11} A^2 = g_{21} A^2 = \frac{3(B^2-1+m)\beta^2}{2}\,.
\eea
Thus $g_{12}, g_{22} > (<)$ 0 provided $B^2 > (<)$ 1 while $g_{11}, 
g_{21}  > (<)$ 0 provided $B^2 < (>)$ $1-m$. On the other hand $\omega_1, 
\omega_2 < 0$. Here $m$ is the modulus of the Jacobi elliptic 
functions \cite{as}. It may be noted that $B > 0$ is true for all the 
thirteen solutions given below.

It is worth pointing out that the local coupled NLS Eqs. (\ref{6}) and 
(\ref{7}) also admit the solution (\ref{8}) provided relations as given by
Eq. (\ref{9}) are satisfied except that $g_{11}$ and $g_{21}$ have opposite 
values compared to those given in (\ref{9}).

{\bf Solution II}

In the limit $m = 1$, the solution I goes over to the hyperbolic solution
\be\label{11}
u(x,t) = e^{i\omega_1 t} \frac{A\tanh(\beta x)}{B+\sech(\beta x)}\,,~~
v(x,t) = e^{i\omega_2 t} \frac{D\sech(\beta x)}{B+\sech(\beta x)}\,,
\ee
provided
\bea\label{12}
&&\omega_1 = \omega_2 = -\frac{\beta^2}{2}\,,~~g_{12} D^2 = 
3 g_{22} D^2 = \frac{3(B^2-1)\beta^2}{2}\,, \nonumber \\
&&B > 0\,,~~3 g_{11} A^2 = g_{21} A^2 = \frac{3 B^2\beta^2}{2}\,. 
\eea
Thus $g_{12}, g_{22} > (<)$ 0 provided $B^2 > (<)$ 1 while $g_{11}, g_{21} 
> 0$. On the other hand $\omega_1, \omega_2 < 0$.

It is worth pointing out that the local coupled NLS Eqs. (\ref{6}) and 
(\ref{7}) also admit the hyperbolic solution (\ref{11}) provided relations 
as given by Eq. (\ref{12}) are satisfied except that $g_{11}$ and $g_{21}$ have 
opposite values compared to those given in (\ref{12}).

{\bf Solution III}

It is easy to check that
\be\label{14}
u(x,t) = e^{i\omega_1 t} \frac{A\sn(\beta x,m)}{B+\cn(\beta x, m)}\,,~~
v(x,t) = e^{i\omega_2 t} \frac{D\dn(\beta x,m)}{B+\cn(\beta x, m)}\,,
\ee
is an exact solution of the coupled Eqs. (\ref{1}) and (\ref{2}) provided
\bea\label{15}
&&\omega_1 = \omega_2 = -\frac{(2m-1)\beta^2}{2}\,,~~g_{12} D^2 = 
3 g_{22} D^2 = \frac{3(B^2-1)\beta^2}{2}\,, \nonumber \\
&&B > 1\,,~~3 g_{11} A^2 = g_{21} A^2 = \frac{3(m B^2+1-m)\beta^2}{2}\,.
\eea
Thus $g_{12}, g_{22}, g_{11}, g_{21}$ are all $ > 0$.
On the other hand $\omega_1, \omega_2 < 0 (>)$ 0 depending on if 
$m > (<)$ 1/2.

It is worth pointing out that the local coupled NLS Eqs. (\ref{6}) and 
(\ref{7}) also admit the solution (\ref{14}) provided relations as given by
Eq. (\ref{15}) are satisfied except that $g_{11}$ and $g_{21}$ have opposite 
values compared to those given in (\ref{9}). 

{\bf Solution IV}

It is easy to check that 
\be\label{16}
u(x,t) = e^{i\omega_1 t} \frac{A}{B+\cos(\beta x)}\,,~~
v(x,t) = e^{i\omega_2 t} \frac{D\sin(\beta x)}{B+\cos(\beta x)}\,,
\ee
with $B > 1$, is an exact solution of the coupled 
Eqs. (\ref{1}) and (\ref{2}) provided
\bea\label{17}
&&\omega_1 = \omega_2 = \frac{\beta^2}{2}\,,~~3 g_{11} A^2 = g_{21} A^2
= 3(B^2-1)\frac{\beta^2}{2}\,, \nonumber \\
&& g_{12} D^2 = 3 g_{22} D^2 = \frac{3\beta^2}{2}\,.
\eea

It is worth pointing out that the local coupled NLS Eqs. (\ref{6}) and 
(\ref{7}) also admit the solution (\ref{23}) provided relations as given by
Eq. (\ref{17}) are satisfied except that $g_{12}$ and $g_{22}$ have opposite 
values compared to those given in (\ref{17}).

{\bf Solution V}

It is straightforward to check that
\be\label{18}
u(x,t) = e^{i\omega_1 t} \frac{A\cos(\beta x)}{1+B\cos^2(\beta x)}\,,~~
v(x,t) = e^{i\omega_2 t} \frac{D\sin(\beta x)}{1+B\cos^2(\beta x)}\,,~~B > 0\,,
\ee
is an exact solution of the nonlocal coupled NLS Eqs. (\ref{1}) and (\ref{2})
provided
\be\label{19}
\omega_1 = \omega_2 = -\beta^2\,,~~g_{12} D^2 = 3 g_{22} D^2 = -6B\beta^2\,,
~~g_{21} A^2 = 3 g_{11} A^2 = -6B(B+1) \beta^2\,.
\ee
Note that the condition $B > 0$ is also valid for the solutions VI to XIII 
presented below and we will not mention it again.

It is worth noting that the solution (\ref{18}) is also a solution of the
coupled local NLS Eqs. (\ref{6}) and (\ref{7})
except that the signs of $g_{12}$ and $g_{22}$ are opposite to those given in
Eq. (\ref{19}).

{\bf Solution VI}

It is easy to check that
\be\label{20}
u(x,t) = e^{i\omega_1 t} \frac{A\cos(\beta x)}{1+B\cos^2(\beta x)}\,,~~
v(x,t) = e^{i\omega_2 t} \frac{D}{1+B\cos^2(\beta x)}\,,
\ee
is an exact solution of the nonlocal coupled Eqs. (\ref{1}) and (\ref{2}) 
provided
\bea\label{21}
&&\omega_1 = -\beta^2\,,~~\omega_2 = -4\beta^2\,,~~g_{12} D^2 = -6B\beta^2\,,
~~g_{11} A^2 = -2B(B+4)\beta^2\,,  \nonumber \\
&&g_{22} D^2 = -2(2-B)\beta^2\,,~~g_{21} A^2 = -6B(B+2) \beta^2\,.
\eea
It is worth noting that the solution (\ref{20}) is also a solution of the
coupled local NLS Eqs. (\ref{6}) and (\ref{7}). 

{\bf Solution VII}

It is easy to check that
\be\label{22}
u(x,t) = e^{i\omega_1 t} \frac{A}{1+B\cos^2(\beta x)}\,,~~
v(x,t) = e^{i\omega_2 t} \frac{D\sin(\beta x)}{1+B\cos^2(\beta x)}\,,
\ee
is an exact solution of the nonlocal coupled Eqs. (\ref{1}) and (\ref{2}) 
provided
\bea\label{23}
&&\omega_1 = -4\beta^2\,,~~g_{12} D^2 =  -6B(B+2)\beta^2\,,
g_{11} A^2 = -2(B+1)(3B+2) \beta^2\,, \nonumber \\
&&\omega_2 = -\beta^2\,,~~g_{22} D^2 = -2B(3B+4)\beta^2\,,
~~g_{21} A^2 = -6B(B+1)\beta^2\,.
\eea
It is worth noting that the solution (\ref{22}) is also a solution of the
coupled local NLS Eqs. (\ref{6}) and (\ref{7}) 
except that the signs of $g_{12}$ and $g_{22}$ are opposite to those given in
Eq. (\ref{23}).

{\bf Solution VIII}

It is easy to check that
\be\label{24}
u(x,t) = e^{i\omega_1 t} \frac{A}{1+B\cos^2(\beta x)}\,,~~v(x,t) 
= e^{i\omega_2 t} \frac{D\sin(\beta x)\cos(\beta x)}{1+B\cos^2(\beta x)}\,,
\ee
is an exact solution of the nonlocal coupled Eqs. (\ref{1}) and (\ref{2}) 
provided
\bea\label{25}
&&\omega_1 = \omega_2 = 2\beta^2\,,~~g_{12} D^2 = 3 g_{22} D^2 =
6B^2\beta^2\,, \nonumber \\
&&g_{21} A^2 = 3 g_{11} A^2 = 6(B+1) \beta^2\,.
\eea
It is worth noting that the solution (\ref{24}) is also a solution of the
coupled local NLS Eqs. (\ref{6}) and (\ref{7}) 
except that the signs of $g_{12}$ and $g_{22}$ are opposite to those given in
Eq. (\ref{25}).

{\bf Solution IX}

It is not difficult to check that
\be\label{26}
u(x,t) = e^{i\omega_1 t} \frac{A\cos(\beta x)}{1+B\cos^2(\beta x)}\,,~~v(x,t) 
= e^{i\omega_2 t} \frac{D\sin(\beta x)\cos(\beta x)}{1+B\cos^2(\beta x)}\,,
\ee
is an exact solution of the nonlocal coupled Eqs. (\ref{1}) and (\ref{2}) 
provided
\bea\label{27}
&&\omega_1 = -6(B+1)\beta^2\,,~~g_{12} D^2 = -6B^3 \beta^2\,,
\nonumber \\
&&g_{11} A^2 = -2B(B+1)(3B+4) \beta^2\,,~~ \omega_2 = -6(B+4) \beta^2\,,
\nonumber \\
&&g_{21} A^2 = -6B(B+1)(B+2) \beta^2\,,~~g_{22} D^2  = -2(3B+2) \beta^2\,.  
\eea
It is worth noting that the solution (\ref{26}) is also a solution of the
coupled local NLS Eqs. (\ref{6}) and (\ref{7}) 
except the signs of $g_{12}$ and $g_{22}$ are opposite to those given in
Eq. (\ref{27}).

{\bf Solution X}

It is easy to check that
\be\label{28}
u(x,t) = \frac{A\sin(\beta x)}{1+B\cos^2(\beta x)}\,,~~
v(x,t) = \frac{D\sin(\beta x)\cos(\beta x)}{1+B\cos^2(\beta x)}\,,
\ee
is an exact solution of coupled Eqs. (\ref{1}) and (\ref{2}) provided
\bea\label{29}
&&(1+B)a_1 = (5B-1)\beta^2\,,~~(1+B)d_1 D^2 =  6B^3 \beta^2\,,
\nonumber \\
&&(1+B) b_1 A^2 = -2B(B+4)\beta^2\,,~~(1+B)a_2 = 2(B-2)\beta^2\,,
\nonumber \\
&&(1+B)d_2 D^2 =  2B^2(B-2) \beta^2\,,~~(1+B) b_2 A^2 = -6B(B+2)\beta^2\,. 
\eea
It is worth noting that the solution (\ref{28}) is also a solution of the
coupled local NLS Eqs. (\ref{6}) and (\ref{7}) 
except the signs of $g_{11}, g_{12}, g_{21}$ and $g_{22}$ are all opposite to 
those given in Eq. (\ref{29}).

{\bf Solution XI}

We now present three soliton solutions with a power law tail. It is easy to 
check that 
\be\label{30}
u(x,t) = e^{i\omega_1 t} \frac{A(F+x^2)}{B+x^2}\,,~~
v(x,t) = e^{i\omega_2 t} \frac{D x}{B+ x^2}\,,
\ee
is an exact solution of the coupled Eqs. (\ref{1}) and (\ref{2}) provided
\bea\label{31}
&&\omega_1 = \omega_2 < 0\,,~~3g_{11} = 4g_{21} < 0\,,~~ 5g_{12} 
= 6g_{22} > 0\,,~~\omega_1  = -\frac{9}{B}\,, \nonumber \\
&&A^2 = \frac{|\omega_1|}{|g_{11}|}\,,~~6d_2 = 5d_1\,,~~
D^2 = \frac{24}{g_{12}}\,,~~F = -B/3\,.
\eea

It is worth pointing out that the local coupled NLS Eqs. (\ref{6}) and 
(\ref{7}) also admit the solution (\ref{30}) provided the relations as given by
Eq. (\ref{31}) are satisfied except that $g_{12}$ and $g_{22}$ have opposite 
values compared to those given in (\ref{31}).

{\bf Solution XII}

It is easy to check that 
\be\label{32}
u(x,t) = e^{i\omega_1 t} \frac{A(F+x^2)}{B+x^2}\,,~~
v(x,t) = e^{i\omega_2 t} \frac{D x}{\sqrt{B+ x^2}}\,,
\ee
is an exact solution of the coupled 
Eqs. (\ref{1}) and (\ref{2}) provided
\bea\label{33}
&&\omega_1 = 8\omega_2 < 0\,,~~3g_{11} = 8g_{21} > 0\,,~~ 7 g_{12} 
= 32 g_{22} > 0\,,~~\omega_1 = -\frac{15}{2B}\,, \nonumber \\
&&5 A^2 = \frac{3|\omega_1|}{|g_{11}|}\,,~~D^2 = \frac{12}{B g_{12}}\,,
~~F = -B/3\,.
\eea

It is worth pointing out that the local coupled NLS Eqs. (\ref{6}) and 
(\ref{7}) also admit the solution (\ref{32}) provided the relations as given by
Eq. (\ref{33}) are satisfied except that $g_{12}$ and $g_{22}$ have opposite 
values compared to those given in (\ref{33}).

{\bf Solution XIII}

It is easy to check that 
\be\label{34}
u(x,t) = e^{i\omega_1 t} \frac{Ax}{B+x^2}\,,~~
v(x,t) = e^{i\omega_2 t} \frac{D x}{\sqrt{B+ x^2}}\,,
\ee
is an exact solution of the coupled 
Eqs. (\ref{1}) and (\ref{2}) provided
\bea\label{35}
&&\omega_1 = 2 \omega_2 < 0\,,~~3g_{11} = 8g_{21} > 0\,,~~7g_{12} 
= 32g_{22} > 0\,,~~\omega_1 = -\frac{6}{B}\,, \nonumber \\
&&g_{11} A^2 = 8\,,~~D^2 = \frac{6}{B g_{12}}\,.
\eea

It is worth pointing out that the local coupled NLS Eqs. (\ref{6}) and 
(\ref{7}) also admit the solution (\ref{34}) provided the relations as given by
Eq. (\ref{35}) are satisfied except that $g_{11},g_{12},g_{21}$ and $g_{22}$ 
have opposite values compared to those given in (\ref{35}).

\subsection{Nonreciprocal Solutions of The Model}

We now show that the coupled nonlocal NLS Eqs. (\ref{1}) and (\ref{2})
admit rather novel {\it nonreciprocal} solutions. Without any loss of 
generality, if we
always choose $u(x,t)$ to be a Lam\'e (inverse Lam\'e) polynomial of 
order two and $v(x,t)$ to be a Lam\'e (inverse Lam\'e) polynomial of
order one, then nonreciprocal solutions exist provided $g_{11} = g_{21} 
=0$. Thus in this case one is effectively solving simpler coupled equations
\be\label{8.1}
iu_{t}(x,t)+u_{xx}(x,t)+g_{12} v(x,t)v^{*}(-x,t)u(x,t) = 0\,,
\ee
\be\label{8.2}
iv_{t}(x,t)+v_{xx}(x,t)+ g_{22} v^2(x,t)v^{*}(-x,t) = 0\,.
\ee
Notice that while $v(x,t)$ satisfies an uncoupled Eq. (\ref{8.2}), 
on the other hand $u(x,t)$ satisfies the coupled Eq. (\ref{8.1}). 
It is worth remembering that if instead we consider the full coupled Eqs. 
(\ref{1}) and (\ref{2}), then u(x,t) being Lam\'e (inverse Lam\'e)  
polynomial of order two and $v(x,t)$ being Lam\'e (inverse Lam\'e) 
polynomial of order one is {\it not} a solution of these coupled equations.
On the other hand, if both $u(x,t)$ and $v(x,t)$ are Lam\'e 
(inverse Lam\'e) polynomials of order one and are solutions of the full
coupled Eqs. (\ref{1}) and (\ref{2}), then they are also the solutions of 
the truncated coupled Eqs. (\ref{8.1}) and (\ref{8.2}).
Hence we will only consider those reciprocal solutions
where $u$ is in terms of Lam\'e (inverse Lam\'e) polynomial of order 
two while $v$ is in terms of Lam\'e (inverse Lam\'e) polynomial of 
order one. We first present twelve nonreciprocal periodic solutions and 
the corresponding six hyperbolic nonreciprocal solutions in terms 
of Lam\'e polynomials of order two in $u$ and order one in $v$. 
Later we will present 12 periodic solutions
in terms of inverse Lam\'e polynomials of order two in $u$ and inverse 
Lam\'e polynomials of order one in $v$.

{\bf Solution XIV}

It is not difficult to check that
\be\label{8.3}
u(x,t) = Ae^{i\omega_1 t} \sqrt{m} \sn(\beta x,m) \dn(\beta x,m)\,,~~
v(x,t) = B e^{i\omega_2 t} \sqrt{m}\sn(\beta x,m)\,,
\ee
is an exact solution of coupled Eqs. (\ref{8.1}) and (\ref{8.2}) provided
\be\label{8.4}
g_{12} B^2 = \beta^2\,,~~3 g_{22} B^2 = \beta^2\,,
\ee
\be\label{8.5}
\omega_1 = -(4m+1) \beta^2\,,~~\omega_2 = -(1+m)\beta^2\,.
\ee
What is remarkable about this solution is that it is independent of the
value of $A$. It is also independent of whether the $u$ field is odd or even
under parity-time reversal ($PT$) symmetry. In fact this feature is common 
to all the reciprocal solutions presented below and hence will not be repeated.

It is worth pointing out that the solution (\ref{8.3}) is also a solution
of the local NLS Eqs. 
\be\label{8.1a}
i u_t(x,t)+u_{xx}(x,t) +6g_{12} |v(x,t)|^2  u(x,t) = 0\,,
\ee
\be\label{8.2a} 
i v_t(x,t)+v_{xx}(x,t) +6g_{22} |v(x,t)|^2 v(x,t) = 0\,,
\ee
except that the signs of $g_{12}$ and $g_{22}$ are opposite compared to
those given in Eq. (\ref{8.4}). 

{\bf Solution XV}

It is easy to check that
\be\label{8.6}
u(x,t) = Ae^{i\omega_1 t} m \sn(\beta x,m) \cn(\beta x,m)\,,~~
v(x,t) = B e^{i\omega_2 t} \sqrt{m}\sn(\beta x,m)\,,
\ee
is an exact solution of coupled Eqs. (\ref{8.1}) and (\ref{8.2}) provided
relations (\ref{8.4}) are satisfied and further
\be\label{8.7}
~\omega_1 = -(4+m)\beta^2\,,~~~\omega_2 = -(1+m)\beta^2\,. 
\ee
It is worth pointing out that the solution (\ref{8.6}) is also a solution
of the local NLS Eqs. (\ref{8.1a}) and (\ref{8.2a})  
except that the signs of $g_{12}$ and $g_{22}$ are opposite compared to
those given in Eq. (\ref{8.4}).

{\bf Solution XVI}

In the limit $m = 1$, both the solutions XIV and XV go over to the hyperbolic
nonreciprocal solution
\be\label{8.8}
u(x,t) = Ae^{i\omega_1 t} \sech(\beta x) \tanh(\beta x)\,,~~
v(x,t) = B e^{i\omega_2 t} \tanh(\beta x)\,,
\ee
provided relations (\ref{8.4}) are satisfied and further
\be\label{8.9}
~\omega_1 = -5\beta^2\,,~~~\omega_2 = -2\beta^2\,. 
\ee
It is worth pointing out that the solution (\ref{8.8}) is also a solution
of the local NLS Eqs. (\ref{8.1a}) and (\ref{8.2a})  
except that the signs of $g_{12}$ and $g_{22}$ are opposite compared to
those given in Eq. (\ref{8.4}).

{\bf Solution XVII}

It is straightforward to check that
\be\label{8.10}
u(x,t) = Ae^{i\omega_1 t} \sqrt{m} \sn(\beta x,m) \dn(\beta x,m)\,,~~
v(x,t) = B e^{i\omega_2 t} \sqrt{m}\cn(\beta x,m)\,,
\ee
is an exact solution of coupled Eqs. (\ref{8.1}) and (\ref{8.2}) provided
relations (\ref{8.4}) are satisfied and further 
\be\label{8.12}
\omega_1 = \omega_2 = (2m-1) \beta^2\,.
\ee

It is worth pointing out that the solution (\ref{8.10}) is also a solution
of the local NLS Eqs. (\ref{8.1a}) and (\ref{8.2a}).

{\bf Solution XVIII}

It is not difficult to check that
\be\label{8.13}
u(x,t) = Ae^{i\omega_1 t} \sqrt{m} \sn(\beta x,m) \dn(\beta x,m)\,,~~
v(x,t) = B e^{i\omega_2 t} \dn(\beta x,m)\,,
\ee
is an exact solution of coupled Eqs. (\ref{8.1}) and (\ref{8.2}) provided
the relations (\ref{8.4}) are satisfied and further
\be\label{8.14}
\omega_1 = (5-4m)\beta^2\,,~~\omega_2 = (2-m) \beta^2\,.
\ee

It is worth pointing out that the solution (\ref{8.13}) is also a solution
of the local NLS Eqs. (\ref{8.1a}) and (\ref{8.2a}).

{\bf Solution XIX}

It is easy to check that
\be\label{8.15}
u(x,t) = Ae^{i\omega_1 t} m \sn(\beta x,m) \cn(\beta x,m)\,,~~
v(x,t) = B e^{i\omega_2 t} \sqrt{m}\cn(\beta x,m)\,,
\ee
is an exact solution of coupled Eqs. (\ref{8.1}) and (\ref{8.2}) provided
the relations (\ref{8.4}) are satisfied and further
\be\label{8.16}
\omega_1 = (5m-4) \beta^2\,,~~\omega_2 = (2m-1) \beta^2\,.
\ee

It is worth pointing out that the solution (\ref{8.15}) is also a solution
of the local NLS Eqs. (\ref{8.1a}) and (\ref{8.2a}).

{\bf Solution XX}

It is easy to check that
\be\label{8.17}
u(x,t) = Ae^{i\omega_1 t} m \sn(\beta x,m) \cn(\beta x,m)\,,~~
v(x,t) = B e^{i\omega_2 t} \dn(\beta x,m)\,,
\ee
is an exact solution of coupled Eqs. (\ref{8.1}) and (\ref{8.2}) provided
the relations (\ref{8.4}) are satisfied and further
\be\label{8.18}
\omega_1 = \omega_2 = (2-m) \beta^2\,.
\ee

It is worth pointing out that the solution (\ref{8.17}) is also a solution
of the local NLS Eqs. (\ref{8.1a}) and (\ref{8.2a}).

{\bf Solution XXI}

In the limit $m = 1$, all the four solutions XVII to XX go over to the
hyperbolic solution
\be\label{8.19}
u(x,t) = Ae^{i\omega_1 t} \sech(\beta x) \tanh(\beta x)\,,~~
v(x,t) = B e^{i\omega_2 t} \sech(\beta x)\,,
\ee
provided the relations (\ref{8.4}) are satisfied and further
\be\label{8.20}
\omega_1 = \omega_2 = \beta^2\,.
\ee

It is worth pointing out that the solution (\ref{8.19}) is also a solution
of the local NLS Eqs. (\ref{8.1a}) and (\ref{8.2a}).

{\bf Solution XXII}

It is easy to check that
\be\label{8.21}
u(x,t) = Ae^{i\omega_1 t} \sqrt{m} \cn(\beta x,m) \dn(\beta x,m)\,,~~
v(x,t) = B e^{i\omega_2 t} \sqrt{m} \sn(\beta x,m)\,,
\ee
is an exact solution of coupled Eqs. (\ref{8.1}) and (\ref{8.2}) provided
the relations (\ref{8.4}) are satisfied and further
\be\label{8.22}
\omega_1 = \omega_2 = -(1+m) \beta^2\,.
\ee

It is worth pointing out that the solution (\ref{8.21}) is also a solution
of the local NLS Eqs. (\ref{8.1a}) and (\ref{8.2a})  
except that the signs of $g_{12}$ and $g_{22}$ are opposite compared to
those given in Eq. (\ref{8.4}).

{\bf Solution XXIII}

It is easy to check that
\be\label{8.23}
u(x,t) = A e^{i\omega_1 t}[\dn^2(\beta x,m)+y]\,,~~
v(x,t) = B e^{i\omega_2 t} \sqrt{m}\sn(\beta x,m)\,,
\ee
is an exact solution of coupled Eqs. (\ref{8.1}) and (\ref{8.2}) provided
relations (\ref{8.4}) are satisfied and further
\be\label{8.24}
\omega_1 = 2(3y+1-2m)\beta^2\,,~~\omega_2 = -(1+m) \beta^2\,,
\ee
where
\be\label{8.25}
y = \frac{-(2-m) \pm \sqrt{1-m+m^2}}{3}\,.
\ee

It is worth pointing out that the solution (\ref{8.23}) is also a solution
of the local NLS Eqs. (\ref{8.1a}) and (\ref{8.2a})  
except that the signs of $g_{12}$ and $g_{22}$ are opposite compared to
those given in Eq. (\ref{8.4}).

{\bf Solution XXIV}

In the limit $m =1$, $y = 0$ or $y = -2/3$. In case $y = 0$ and $m = 1$, both 
the solutions XXII and XXIII go over to the hyperbolic solution 
\be\label{8.26}
u(x,t) = A e^{i\omega_1 t} \sech^2(\beta x)\,,~~
v(x,t) = B\sqrt{m}\tanh(\beta x)\,,
\ee
provided relations (\ref{8.4}) are satisfied and further
\be\label{8.27}
\omega_1 = \omega_2 = -2 \beta^2\,.
\ee

It is worth pointing out that the solution (\ref{8.26}) is also a solution
of the local NLS Eqs. (\ref{8.1a}) and (\ref{8.2a})  
except that the signs of $g_{12}$ and $g_{22}$ are opposite compared to
those given in Eq. (\ref{8.4}).

{\bf Solution XXV}

On the other hand, in case $y = -2/3$ and $m = 1$, the 
solution XXIII goes over to the hyperbolic solution 
\be\label{8.28}
u(x,t) =  A e^{i\omega_1 t} [\sech^2(\beta x)-2/3]\,,~~
v(x,t) = B\sqrt{m}\tanh(\beta x)\,,
\ee
provided relations (\ref{8.4}) are satisfied and further
\be\label{8.29}
\omega_1 = -6\beta^2\,,~~\omega_2 = -2 \beta^2\,.
\ee

It is worth pointing out that the solution (\ref{8.28}) is also a solution
of the local NLS Eqs. (\ref{8.1a}) and (\ref{8.2a})  
except that the signs of $g_{12}$ and $g_{22}$ are opposite compared to
those given in Eq. (\ref{8.4}).

{\bf Solution XXVI}

It is easy to check that
\be\label{8.30}
u(x,t) = Ae^{i\omega_1 t} \sqrt{m} \cn(\beta x,m) \dn(\beta x,m)\,,~~
v(x,t) = B e^{i\omega_2 t} \sqrt{m} \cn(\beta x,m)\,,
\ee
is an exact solution of coupled Eqs. (\ref{8.1}) and (\ref{8.2}) provided
the relations (\ref{8.4}) are satisfied and further
\be\label{8.31}
\omega_1 = (5m-1)\beta^2\,,~~\omega_2 = (2m-1) \beta^2\,.
\ee

It is worth pointing out that the solution (\ref{8.30}) is also a solution
of the local NLS Eqs. (\ref{8.1a}) and (\ref{8.2a}).

{\bf Solution XXVII}

It is straightforward to check that
\be\label{8.32}
u(x,t) = Ae^{i\omega_1 t} \sqrt{m} \cn(\beta x,m) \dn(\beta x,m)\,,~~
v(x,t) = B e^{i\omega_2 t} \dn(\beta x,m)\,,
\ee
is an exact solution of coupled Eqs. (\ref{8.1}) and (\ref{8.2}) provided
the relations (\ref{8.4}) are satisfied and further
\be\label{8.33}
\omega_1 = (5-m)\beta^2\,,~~\omega_2 = (2-m) \beta^2\,.
\ee

It is worth pointing out that the solution (\ref{8.32}) is also a solution
of the local NLS Eqs. (\ref{8.1a}) and (\ref{8.2a}).

{\bf Solution XXVIII}

It is easy to check that
\be\label{8.34}
u(x,t) = A e^{i\omega_1 t} [\dn^2(\beta x,m)+y]\,,~~
v(x,t) = B e^{i\omega_2 t} \sqrt{m}\cn(\beta x,m)\,,
\ee
is an exact solution of coupled Eqs. (\ref{8.1}) and (\ref{8.2}) provided
the relations (\ref{8.4}) are satisfied and further
\be\label{8.35}
\omega_1 = 2[(1+m)+3y]\beta^2\,,~~\omega_2 = (2m-1) \beta^2\,,
\ee
where $y$ is again as given by Eq. (\ref{8.25}).

It is worth pointing out that the solution (\ref{8.34}) is also a solution
of the local NLS Eqs. (\ref{8.1a}) and (\ref{8.2a}).

{\bf Solution XXIX}

It is easy to check that
\be\label{8.36}
u(x,t) = A e^{i\omega_1 t} [\dn^2(\beta x,m)+y]\,,~~
v(x,t) = B e^{i\omega_2 t} \dn(\beta x,m)\,,
\ee
is an exact solution of coupled Eqs. (\ref{8.1}) and (\ref{8.2}) provided
the relations (\ref{8.4}) are satisfied and further
\be\label{8.37}
\omega_1 = 2[2(2-m)+3y]\beta^2\,,~~\omega_2 = (2-m) \beta^2\,,
\ee
where $y$ is again as given by Eq. (\ref{8.25}).

It is worth pointing out that the solution (\ref{8.36}) is also a solution
of the local NLS Eqs. (\ref{8.1a}) and (\ref{8.2a}).

{\bf Solution XXX}

In the limit $m =1$, $y = 0$ or $y = -2/3$. In case $y = 0$ and $m = 1$, all 
the four solutions XXVI to XXIX go over to the hyperbolic solution 
\be\label{8.38}
u(x,t) = A e^{i\omega_1 t} \sech^2(\beta x)\,,~~
v(x,t) = B e^{i\omega_2 t} \sech(\beta x)\,,
\ee
provided relations (\ref{8.4}) are satisfied and further
\be\label{8.39}
\omega_1 = 4\beta^2\,,~~\omega_2 = \beta^2\,.
\ee

It is worth pointing out that the solution (\ref{8.38}) is also a solution
of the local NLS Eqs. (\ref{8.1a}) and (\ref{8.2a}).

{\bf Solution XXXI}

On the other hand, in case $y = -2/3$ and $m = 1$, the solutions XXVIII and 
XXIX go over to the solution
\be\label{8.40}
u(x,t) = A e^{i\omega_1 t} [\sech^2(\beta x)-2/3]\,,~~
v(x,t) = B e^{i\omega_2 t} \sech(\beta x)\,,
\ee
provided relations (\ref{8.4}) are satisfied and further
\be\label{8.41}
\omega_1 = 0\,,~~\omega_2 = \beta^2\,.
\ee

It is worth pointing out that the solution (\ref{8.40}) is also a solution
of the local NLS Eqs. (\ref{8.1a}) and (\ref{8.2a}).

I now present two novel superposed solutions.

{\bf Solution XXXII}

It is easy to check that
\be\label{8.42}
u(x,t) =  e^{i\omega_1 t} [A\dn^2(\beta x, m)+ Ay +B\sqrt{m} \cn(\beta x, m) 
\dn(\beta x, m)]\,,
\ee
\be\label{8.43}
v(x,t) = D e^{i\omega_2 t} [\dn(\beta x, m)+E\sqrt{m} \cn(\beta x, m)]\,,
\ee
is an exact solution of coupled Eqs. (\ref{8.1}) and (\ref{8.2}) provided
\bea\label{8.44}
&&4 g_{12} D^2 = \beta^2\,,~~\omega_1 = [\frac{(7+m)}{2}+3y]\beta^2\,,~~
E = \pm D\,,~~B = \pm A\,, \nonumber \\
&&\omega_2 = \frac{(1+m)\beta^2}{2}\,,~~12 g_{22} D^2 = \beta^2\,, \nonumber \\
&&y = \frac{-(5-m) \pm \sqrt{1+14m+m^2}}{6}\,.
\eea
Note that the $\pm$ signs in $D, E$ are correlated with the signs in 
$B$ and $A$. Further, the solution is independent of the values of $B$ and $A$ 
except that they are constrained by $B = A$. In the limit $m = 1$, this 
solution goes over to the solution XXX.

It is worth pointing out that the solution (\ref{8.42}) is also a solution
of the local NLS Eqs. (\ref{8.1a}) and (\ref{8.2a}).

{\bf Solution XXXIII}

It is easy to check that
\be\label{8.45}
u(x,t) = A e^{i\omega_1 t} [\sqrt{m}\dn(\beta x, m) \sn(\beta x, m)
+B m \cn(\beta x, m) \sn(\beta x, m)]\,,
\ee
\be\label{8.46}
v(x,t) = e^{i\omega_2 t} [D\dn(\beta x, m)+E\sqrt{m} \cn(\beta x, m)]\,,
\ee
is an exact solution of coupled Eqs. (\ref{8.1}) and (\ref{8.2}) provided
\bea\label{8.47}
&&4 g_{12} D^2 = \beta^2\,,~~\omega_1 = \omega_2 = \frac{(1+m)\beta^2}{2}\,,~~
E = \pm D\,,~~B = \pm A\,, \nonumber \\
&&12 g_{22} D^2 = \beta^2\,.
\eea
Note that the $\pm$ signs in $D, E$ are correlated with the signs in 
$B$ and $A$. Further, the solution is independent of the values of $A$ and 
$B$ except that they must satisfy the constraint $B = \pm A$. In the limit 
$m = 1$, this solution goes over to the hyperbolic solution XXI.

It is worth pointing out that the solution (\ref{8.45}) is also a solution
of the local NLS Eqs. (\ref{8.1a}) and (\ref{8.2a}) except that the signs
of $g_{12}$ and $g_{22}$ are opposite to those given in Eq. (\ref{8.47}).

Remarkably, these coupled equations also admit several solutions in terms of 
inverse Lam\'e polynomial solutions of order two in $u(x,t)$ and of inverse
Lam\'e polynomial solutions of order one in $v(x,t)$. We list these solutions one by 
one. It may be noted that all these solutions are only valid for $0 < m < 1$.

{\bf Solution XXXIV}

It is easy to check that
\be\label{9.1}
u(x,t) = e^{i\omega_1 t} \frac{A\sqrt{m}\cn(\beta x,m)}{\dn^2(\beta x,m)}\,,~~
v(x,t) = e^{i\omega_2 t} \frac{B\sqrt{m}\cn(\beta x,m)}{\dn(\beta x,m)}\,,
\ee
is an exact solution of coupled Eqs. (\ref{8.1}) and (\ref{8.2}) provided
\be\label{9.2}
g_{12} B^2 = -6\beta^2\,,~~g_{22} B^2 = -2\beta^2\,, 
\ee
\be\label{9.3}
\omega_1 = -(4m+1)\beta^2\,,~~\omega_2 = -(1+m)\beta^2\,. 
\ee
It is worth pointing out that the solution (\ref{9.1}) is also a solution
of the local NLS Eqs. (\ref{8.1a}) and (\ref{8.2a}). Note that this solution
is independent of the magnitude of $A$. In fact this feature is common to all
the solutions given below.

{\bf Solution XXXV}

It is easy to check that
\be\label{9.4}
u(x,t) = e^{i\omega_1 t} \frac{A\sqrt{m}\cn(\beta x,m)}{\dn^2(\beta x,m)}\,,~~
v(x,t) = e^{i\omega_2 t} \frac{B\sqrt{m}\sn(\beta x,m)}{\dn(\beta x,m)}\,,
\ee
is an exact solution of coupled Eqs. (\ref{8.1}) and (\ref{8.2}) provided
\be\label{9.5}
g_{12} B^2 = -6(1-m)\beta^2\,,~~g_{22} B^2 = -2(1-m) \beta^2\,,
\ee
\be\label{9.6}
\omega_1 = \omega_2 = (2m-1)\beta^2\,.
\ee

It is worth pointing out that the solution (\ref{9.4}) is also a solution
of the local NLS Eqs. (\ref{8.1a}) and (\ref{8.2a}) except that the signs
of $g_{12}$ and $g_{22}$ are opposite to those given in Eq. (\ref{9.5}).

{\bf Solution XXXVI}

It is easy to check that
\be\label{9.7}
u(x,t) = e^{i\omega_1 t} \frac{A\sqrt{m}\cn(\beta x,m)}{\dn^2(\beta x,m)}\,,~~
v(x,t) = e^{i\omega_2 t} \frac{B}{\dn(\beta x,m)}\,,
\ee
is an exact solution of coupled Eqs. (\ref{8.1}) and (\ref{8.2}) provided
\be\label{9.8}
g_{12} B^2 = 6(1-m)\beta^2\,,~~g_{22} B^2 = 2(1-m)\beta^2\,,
\ee
\be\label{9.9}
\omega_1 = (5-4m) \beta^2\,,~~\omega_2 = (2-m)\beta^2\,.
\ee
It is worth pointing out that the solution (\ref{9.7}) is also a solution
of the local NLS Eqs. (\ref{8.1a}) and (\ref{8.2a}).

{\bf Solution XXXVII}

It is easy to check that
\be\label{9.10}
u(x,t) = e^{i\omega_1 t} \frac{A\sqrt{m}\sn(\beta x,m)}{\dn^2(\beta x,m)}\,,~~
v(x,t) = e^{i\omega_2 t} \frac{B\cn(\beta x,m)}{\dn(\beta x,m)}\,,
\ee
is an exact solution of coupled Eqs. (\ref{8.1}) and (\ref{8.2}) provided
\be\label{9.11}
g_{12} B^2 = -6\beta^2\,,~~g_{22} B^2 = -2\beta^2\,,
\ee
\be\label{9.12}
\omega_1 = \omega_2 = -(1+m) \beta^2\,.
\ee

It is worth pointing out that the solution (\ref{9.10}) is also a solution
of the local NLS Eqs. (\ref{8.1a}) and (\ref{8.2a}).

{\bf Solution XXXVIII}

It is easy to check that
\be\label{9.13}
u(x,t) = e^{i\omega_1 t} \frac{A\sqrt{m}\sn(\beta x,m)}{\dn^2(\beta x,m)}\,,~~
v(x,t) = e^{i\omega_2 t} \frac{B\sn(\beta x,m)}{\dn(\beta x,m)}\,,
\ee
is an exact solution of coupled Eqs. (\ref{8.1}) and (\ref{8.2}) provided
\be\label{9.14}
g_{12} B^2 = -6(1-m)\beta^2\,,~~g_{22} B^2 = -2(1-m)\beta^2\,,
\ee
\be\label{9.15}
\omega_1 = (5m-1)\beta^2\,,~~\omega_2 = (2m-1) \beta^2\,.
\ee

It is worth pointing out that the solution (\ref{9.13}) is also a solution
of the local NLS Eqs. (\ref{8.1a}) and (\ref{8.2a}) except that the signs
of $g_{12}$ and $g_{22}$ are opposite to those given in Eq. (\ref{9.14}).

{\bf Solution XXXIX}

It is easy to check that
\be\label{9.16}
u(x,t) = e^{i\omega_1 t} \frac{A\sqrt{m}\sn(\beta x,m)}{\dn^2(\beta x,m)}\,,~~
v(x,t) = e^{i\omega_2 t} \frac{B}{\dn(\beta x,m)}\,,
\ee
is an exact solution of coupled Eqs. (\ref{8.1}) and (\ref{8.2}) provided
\be\label{9.17}
g_{12} B^2 = 6(1-m)\beta^2\,,~~g_{22} B^2 = 2(1-m)\beta^2\,,
\ee
\be\label{9.18}
\omega_1 = -(5-m)\beta^2\,,~~\omega_2 = (2-m) \beta^2\,.
\ee

It is worth pointing out that the solution (\ref{9.16}) is also a solution
of the local NLS Eqs. (\ref{8.1a}) and (\ref{8.2a}).

{\bf Solution XXXX}

It is easy to check that
\be\label{9.19}
u(x,t) = e^{i\omega_1 t} \frac{A\sqrt{m}\sn(\beta x,m)\cn(\beta x, m)}
{\dn^2(\beta x,m)}\,,
\ee
\be\label{9.20}
v(x,t) = e^{i\omega_2 t} \frac{B\sqrt{m} \cn(\beta x,m)}{\dn(\beta x,m)}\,,
\ee
is an exact solution of coupled Eqs. (\ref{8.1}) and (\ref{8.2}) provided
\be\label{9.21}
g_{12} B^2 = -6\beta^2\,,~~g_{22} B^2 = -2\beta^2\,,
\ee
\be\label{9.22}
\omega_1 = -(4+m)\beta^2\,,~~\omega_2 = -(1+m) \beta^2\,.
\ee

It is worth pointing out that the solution (\ref{9.19}) and (\ref{9.20}) 
is also a solution of the local NLS Eqs. (\ref{8.1a}) and (\ref{8.2a}).

{\bf Solution XXXXI}

It is easy to check that
\be\label{9.23}
u(x,t) = e^{i\omega_1 t} \frac{A\sqrt{m}\sn(\beta x,m)\cn(\beta x, m)}
{\dn^2(\beta x,m)}\,,
\ee
\be\label{9.24}
v(x,t) = e^{i\omega_2 t} \frac{B\sqrt{m} \sn(\beta x,m)}{\dn(\beta x,m)}\,,
\ee
is an exact solution of coupled Eqs. (\ref{8.1}) and (\ref{8.2}) provided
\be\label{9.25}
g_{12} B^2 = -6(1-m)\beta^2\,,~~g_{22} B^2 = -2(1-m)\beta^2\,,
\ee
\be\label{9.26a}
\omega_1 = -(5m-4)\beta^2\,,~~\omega_2 = (2m-1) \beta^2\,.
\ee

It is worth pointing out that the solution (\ref{9.23}) and (\ref{9.24}) 
is also a solution of the local NLS Eqs. (\ref{8.1a}) and (\ref{8.2a}) except 
that the signs of $g_{12}$ and $g_{22}$ are opposite to those given in 
Eq. (\ref{9.25}).

{\bf Solution XXXXII}

It is easy to check that
\be\label{9.26}
u(x,t) = e^{i\omega_1 t} \frac{A\sqrt{m}\sn(\beta x,m)\cn(\beta x, m)}
{\dn^2(\beta x,m)}\,,
\ee
\be\label{9.27}
v(x,t) = e^{i\omega_2 t} \frac{B}{\dn(\beta x,m)}\,,
\ee
is an exact solution of coupled Eqs. (\ref{8.1}) and (\ref{8.2}) provided
\be\label{9.28}
g_{12} B^2 = 6(1-m)\beta^2\,,~~g_{22} B^2 = 2(1-m)\beta^2\,,
\ee
\be\label{9.29}
\omega_1 = \omega_2 = (2-m)\beta^2\,.
\ee

It is worth pointing out that the solution (\ref{9.26}) and (\ref{9.27}) 
is also a solution of the local NLS Eqs. (\ref{8.1a}) and (\ref{8.2a}).

{\bf Solution XXXXIII}

It is easy to check that
\be\label{9.30}
u(x,t) = e^{i\omega_1 t} A\left[\frac{1}{\dn^2(\beta x,m)}+y\right]\,,
\ee
\be\label{9.31}
v(x,t) = e^{i\omega_2 t} \frac{B\sqrt{m}\cn(\beta x,m)}{\dn(\beta x,m)}\,,
\ee
is an exact solution of coupled Eqs. (\ref{8.1}) and (\ref{8.2}) provided
\be\label{9.32}
g_{12} B^2 = -6\beta^2\,,~~g_{22} B^2 = -2\beta^2\,,
\ee
\be\label{9.33}
\omega_1 = -2(3+\frac{1}{y})\beta^2\,,~~\omega_2 = (2-m)\beta^2\,,
\ee
where
\be\label{9.34}
y = \frac{-(2-m) \pm \sqrt{1-m+m^2}}{3(1-m)}\,.
\ee

It is worth pointing out that the solution (\ref{9.30}) and (\ref{9.31}) 
is also a solution of the local NLS Eqs. (\ref{8.1a}) and (\ref{8.2a}).

{\bf Solution XXXXIV}

It is not difficult to check that
\be\label{9.35}
u(x,t) = e^{i\omega_1 t} A\left[\frac{1}{\dn^2(\beta x,m)}+y\right]\,,
\ee
\be\label{9.36}
v(x,t) = e^{i\omega_2 t} \frac{B\sqrt{m}\sn(\beta x,m)}{\dn(\beta x,m)}\,,
\ee
is an exact solution of coupled Eqs. (\ref{8.1}) and (\ref{8.2}) provided
\be\label{9.37}
g_{12} B^2 = -6(1-m)\beta^2\,,~~g_{22} B^2 = -2(1-m)\beta^2\,,
\ee
\be\label{9.38}
\omega_1 = -2[3(1-m)+\frac{1}{y}]\beta^2\,,~~\omega_2 = (2m-1)\beta^2\,,
\ee
while $y$ is again as given by Eq. (\ref{9.34}).

It is worth pointing out that the solution (\ref{9.35}) and (\ref{9.36}) 
is also a solution of the local NLS Eqs. (\ref{8.1a}) and (\ref{8.2a}) except 
that the signs of $g_{12}$ and $g_{22}$ are opposite to those given in 
Eq. (\ref{9.37}).

{\bf Solution XXXXV}

It is easy to check that
\be\label{9.39}
u(x,t) = e^{i\omega_1 t} A\left[\frac{1}{\dn^2(\beta x,m)}+y\right]\,,
\ee
\be\label{9.40}
v(x,t) = e^{i\omega_2 t} \frac{B}{\dn(\beta x,m)}\,,
\ee
is an exact solution of coupled Eqs. (\ref{8.1}) and (\ref{8.2}) provided
\be\label{9.41}
g_{12} B^2 = 6(1-m)\beta^2\,,~~g_{22} B^2 = (2-m)\beta^2\,,
\ee
\be\label{9.42}
\omega_1 = -\frac{2}{y}\beta^2\,,~~\omega_2 = (2-m)\beta^2\,,
\ee
while $y$ is again as given by Eq. (\ref{9.34}).

It is worth pointing out that the solution (\ref{9.39}) and (\ref{9.40}) 
is also a solution of the local NLS Eqs. (\ref{8.1a}) and (\ref{8.2a}).

Finally, we present a novel solution where while $\phi_1$ is a combination
of Lam\'e polynomials of order two and inverse Lam\'e polynomials of order
two,  $\phi_2$ is a combination of Lam\'e polynomials of order one and inverse 
Lam\'e polynomials of order one.

{\bf Solution XXXXVI}

It is easy to check that
\be\label{9.43}
u(x,t) = e^{i\omega_1 t} \bigg (A[\dn^2(\beta x, m)+\sqrt{1-m}y] 
+\frac{B (1-m)}{\dn^2(\beta x, m)} \bigg )\,, 
\ee
\be\label{9.44}
v(x,t) = e^{i\omega_2 t} [D\dn(\beta x, m) +\frac{E\sqrt{(1-m)}}
{\dn(\beta x, m)}]\,,
\ee
is an exact solution of coupled Eqs. (\ref{8.1}) and (\ref{8.2}) provided
\bea\label{9.45}
&&g_{12} D^2 = 6\beta^2\,,~~\omega_2 = (2-m\pm 6\sqrt{1-m})\beta^2\,,~~
E = \pm D\,,~~B = \pm A\,, \nonumber \\
&&g_{22} D^2 = 2\beta^2\,,~~\omega_1 = 2[2(2-m)+3y\pm 6\sqrt{1-m}]\beta^2\,,
\nonumber \\
&&y = \frac{-(2-m) \pm 6\sqrt{1-m}}{3}\,,~~0 < m < 1\,.
\eea
Note that the $\pm$ signs in the various relations are correlated. Further,
the solution is independent of the values of $A$ and $B$ except that they
must satisfy the constraint $B = \pm A$. 

It is worth pointing out that the solution (\ref{9.43}) and (\ref{9.44}) 
is also a solution of the local NLS Eqs. (\ref{8.1a}) and (\ref{8.2a}).

\section{New Solutions of a Coupled Nonlocal mKdV Model}

Let us consider the following nonlocal coupled mKdV model \cite{ks22}
\be\label{1.1}
u_{t}(x,t)+u_{xxx}(x,t) +6[g_{11} u(x,t) u(-x,-t) 
+g_{12} v(x,t) v(-x,-t)]u_{x}(x,t) = 0\,, 
\ee
\be\label{1.2}
v_{t}(x,t)+v_{xxx}(x,t) +6[g_{21} u(x,t) u(-x,-t) 
+g_{22} v(x,t) v(-x,-t)]v_{x}(x,t) = 0\,.
\ee

First of all, notice that unlike the coupled local mKdV, the nonlocal
coupled mKdV Eqs. (\ref{1.1}) and (\ref{1.2}) admit a plane wave solution
\be\label{1.3d}
u(x,t) = A e^{i(k_1 x-\omega_1 t)}\,,~~v(x,t) = B e^{i(k_2 x -\omega_2 t)}\,,
\ee
provided
\be\label{1.3e}
\omega_1 +k_{1}^{3} = 6k_1[g_{11}A^2 +g_{12} B^2]\,,~~
\omega_2 +k_{2}^{3} = 6k_2[g_{21}A^2 +g_{22} B^2]\,.
\ee
It is obvious from here that even the uncoupled nonlocal mKdV equation
\be\label{1.4g}
u_{t}(x,t)+u_{xxx}(x,t) +6 g  u(x,t) u(-x,-t) u_{x}(x,t) = 0\,, 
\ee
admits the plane wave solution
\be\label{1.4h}
u(x,t) = A e^{i(kx -\omega t)}\,,
\ee
provided 
\be\label{1.4j}
\omega_1 +k_{1}^{3} = 6k_1 A^2 \,.
\ee

We now show that unlike the coupled local mKdV equations
\be\label{1.5}
u_{t}(x,t)+u_{xxx}(x,t) +6[g_{11} u^2(x,t) 
+g_{12} v^2(x,t)]u_{x}(x,t) = 0\,,
\ee
\be\label{1.6}
v_{t}(x,t)+v_{xxx}(x,t) +6[g_{21} u^2(x,t) 
+g_{22} v^2(x,t)]v_{x}(x,t) = 0\,,
\ee
the nonlocal mKdV Eqs. (\ref{1.1}) and (\ref{1.2}) admit solutions
similar to those obtained by Zakharav and Shabat \cite{zs} for the
local uncoupled NLS case.

{\bf Solution I}

It is straightforward to check that
\bea\label{2.1}
&&u(x,t) = \sqrt{n_1} [B\tanh(\xi)+iA] e^{i(k_1 x-\omega_1 t)}\,,
\nonumber \\
&&v(x,t) = \sqrt{n_2} [D\tanh(\xi)+iE] e^{i(k_2 x-\omega_2 t)}\,,
~~\xi = \beta(x-ct)\,,
\eea
is an exact solution of the nonlocal coupled Eqs. (\ref{1.1}) and (\ref{1.2}) 
provided
\be\label{2.2}
\beta^2 = g_{11} n_1 B^2 + g_{12} n_2 D^2 = g_{21} n_1 B^2 + g_{22} n_2 
D^2\,, 
\ee
\be\label{2.3}
\omega_1+k_{1}^3 = -6\beta^2 k_1 -6k_1[g_{11} n_1 A^2 +g_{12} n_2 E^2]\,,
\ee
\be\label{2.4}
\omega_2+k_{2}^3 = -6\beta^2 k_2 -6k_2[g_{21} n_1 A^2 +g_{22} n_2 E^2]\,,
\ee
\be\label{2.5}
Bck_1 = -2B k_{1}^3 +4k_1 B \beta^2 -6k_{1}^2 A \beta +B\omega_1\,,
\ee
\be\label{2.6}
Dck_2 = -2D k_{2}^3 +4k_2 D \beta^2 -6k_{2}^2 E \beta +D\omega_2\,.
\ee

{\bf Solution II}

It is easy to check that
\bea\label{2.7}
&&u(x,t) = \sqrt{n_1} [B\tanh(\xi)+iA] e^{i(k_1 x-\omega_1 t)}\,,
\nonumber \\
&&v(x,t) = \sqrt{n_2} [E+iD\tanh(\xi)] e^{i(k_2 x-\omega_2 t)}\,,
\eea
is an exact solution of the nonlocal coupled Eqs. (\ref{1.1}) and (\ref{1.2}) 
provided
\be\label{2.8}
\beta^2 = g_{11} n_1 B^2 - g_{12} n_2 D^2 = g_{21} n_1 B^2 - g_{22} n_2 
D^2\,, 
\ee
\be\label{2.9}
\omega_1+k_{1}^3 = -6\beta^2 k_1 -6k_1[g_{11} n_1 A^2 - g_{12} n_2 E^2]\,,
\ee
\be\label{2.10}
\omega_2+k_{2}^3 = -6\beta^2 k_2 -6k_2[g_{21} n_1 A^2 - g_{22} n_2 E^2]\,,
\ee
\be\label{2.11}
Bck_1 = -2B k_{1}^3 +4k_1 B \beta^2 -6k_{1}^2 A \beta +B\omega_1\,,
\ee
\be\label{2.12}
Dck_2 = -2D k_{2}^3 +4k_2 D \beta^2 +6k_{2}^2 E \beta +D\omega_2\,.
\ee

{\bf Solution III}

It is not difficult to check that
\bea\label{2.13}
&&u(x,t) = \sqrt{n_1} [A+iB\tanh(\xi)] e^{i(k_1 x-\omega_1 t)}\,,
\nonumber \\
&&v(x,t) = \sqrt{n_2} [E+iD\tanh(\xi)] e^{i(k_2 x-\omega_2 t)}\,,
\eea
is an exact solution of the nonlocal coupled Eqs. (\ref{1.1}) and (\ref{1.2}) 
provided
\be\label{2.14}
\beta^2 = -[g_{11} n_1 B^2 + g_{12} n_2 D^2 
= -[g_{21} n_1 B^2 + g_{22} n_2 D^2]\,, 
\ee
\be\label{2.15}
\omega_1+k_{1}^3 = -6\beta^2 k_1 +6k_1[g_{11} n_1 A^2 + g_{12} n_2 E^2]\,,
\ee
\be\label{2.16}
\omega_2+k_{2}^3 = -6\beta^2 k_2 +6k_2[g_{21} n_1 A^2 + g_{22} n_2 E^2]\,,
\ee
\be\label{2.17}
Bck_1 = -2B k_{1}^3 +4k_1 B \beta^2 +6k_{1}^2 A \beta +B\omega_1\,,
\ee
\be\label{2.18}
Dck_2 = -2D k_{2}^3 +4k_2 D \beta^2 +6k_{2}^2 E \beta +D\omega_2\,.
\ee
Note that the coupled local mKdV Eqs. (\ref{1.5}) and (\ref{1.6}) do not
admit the solutions I, II and III.

We now show that corresponding to any solution
of the coupled nonlocal Eqs. (\ref{1.1}) and (\ref{1.2}) of the form 
\be\label{1.3f}
u(x,t) = u(\xi)\,,~~v(x,t) = v(\xi)\,,~~\xi = \beta(x-ct)\,,
\ee
where $u(\xi)$ and $v(\xi)$ are real, there are always solutions of the form 
\be\label{1.3g}
u(x,t) = u(\xi) e^{i(k_1 x-\omega_1 t)}\,,~~v(x,t) 
= v(\xi) e^{i(k_2 x -\omega_2 t)}\,,~~\xi = \beta(x-vt)\,.
\ee
The proof is straightforward. On substituting the ansatz as given by 
Eq. (\ref{1.3g}) in the coupled Eqs. (\ref{1.1}) and (\ref{1.2}) and equating 
the real and imaginary parts we obtain the following four relations
\be\label{1.3h}
(\omega_1+k_{1}^3) u(\xi) = 3\beta^2 k_1 u''(\xi) +6k_1[g_{11} u(\xi)u(-\xi)
+g_{12} v(\xi) v(-\xi)] u(\xi)\,,
\ee
\be\label{1.3i}
(c+3k_{1}^2) u'(\xi) = \beta^2  u'''(\xi) +6[g_{11} u(\xi)u(-\xi)
+g_{12} v(\xi) v(-\xi)]u'(\xi)\,,
\ee
\be\label{1.3j}
(\omega_2+k_{2}^3)v(\xi) = 3\beta^2 k_2 v''(\xi) +6k_2[g_{21} u(\xi)u(-\xi)
+g_{12} v(\xi) v(-\xi)] v(\xi)\,,
\ee
\be\label{1.3k}
(c+3k_{2}^2) v'(\xi) = \beta^2  v'''(\xi) +6[g_{21} u(\xi)u(-\xi)
+g_{22} v(\xi) v(-\xi)]v'(\xi)\,. 
\ee
In the limit $k_1 = k_2 = \omega_1 = \omega_2 = 0$, the two Eqs. (\ref{1.3h})
as well as (\ref{1.3j}) disappear while the other two equations lead to
the two coupled equations for $u(\xi)$ and $v(\xi)$
\be\label{1.3l}
c u'(\xi) = \beta^2  u'''(\xi) +6[g_{11} u(\xi)u(-\xi)
+g_{12} v(\xi) v(-\xi)]u'(\xi)\,,
\ee
\be\label{1.3m}
c v'(\xi) = \beta^2  v'''(\xi) +6[g_{11} u(\xi)u(-\xi)
+g_{12} v(\xi) v(-\xi)]v'(\xi)\,.
\ee
We reemphasize that the corresponding local mKdV coupled Eqs. (\ref{1.5})
and (\ref{1.6}) do not admit either the plane wave type solutions of the form 
(\ref{1.3d}) nor the solutions of the form (\ref{1.3g}). As an illustration, we now 
present six well known solutions of the nonlocal coupled mKdV Eqs. (\ref{1.1}) 
and (\ref{1.2}) without the plane wave factor but reconsider them multiplied by 
the extra plane wave factor.

{\bf Solution IV}

It is not difficult to check that
\be\label{1.2b}
u(x,t) = A \sqrt{m} \sn(\xi,m) e^{i(k_1 x -\omega_1 t)}\,,~~
v(x,t) = B \sqrt{m} \cn(\xi,m) e^{i(k_2 x -\omega_2 t)}\,,
\ee
is an exact solution of the coupled Eqs. (\ref{1.1}) and (\ref{1.2}) 
provided
\bea\label{1.2c}
&&\beta^2 = g_{11} A^2 + g_{12} B^2 = g_{21} A^2 + g_{22} B^2\,,
\nonumber \\
&&c+3k_{1}^{2} = (5m-1)\beta^2 -6m g_{11} A^2\,,~~
\omega_2+k_{2}^3 = 3(2m-1) k_2 \beta^2 -6m k_2 g_{21} A^2\,, 
\nonumber \\
&&c+3k_{2}^{2} = (2m-1)\beta^2 -6m g_{21} A^2\,,~~
\omega_1+k_{1}^3 = -3(1-m) k_1 \beta^2 -6m k_1 g_{11} A^2\,. \nonumber\\  
\eea

{\bf Solution V}

It is easy to check that
\be\label{1.2d}
u(x,t) = A \sqrt{m} \sn(\xi,m) e^{i(k_1 x -\omega_1 t)}\,,~~
v(x,t) = B \dn(\xi,m) e^{i(k_2 x -\omega_2 t)}\,,
\ee
is an exact solution of the coupled Eqs. (\ref{1.1}) and (\ref{1.2}) 
provided
\bea\label{1.2e}
&&\beta^2 = g_{11} A^2 + g_{12} B^2 = g_{21} A^2 + g_{22} B^2\,,
\nonumber \\
&&c+3k_{1}^{2} = (5-m)\beta^2 -6 g_{11} A^2\,,~~
\omega_2+k_{2}^3 = 3(2-m) k_2 \beta^2 -6 k_2 g_{21} A^2\,, 
\nonumber \\
&&c+3k_{2}^{2} = (2-m)\beta^2 -6 g_{21} A^2\,,~~
\omega_1+k_{1}^3 = 3(1-m) k_1 \beta^2 -6 k_1 g_{11} A^2\,. \nonumber\\  
\eea

{\bf Solution VI}

In the limit $m = 1$, both the solutions I and II go over to
the hyperbolic solution
\be\label{1.2f}
u(x,t) = A \tanh(\xi) e^{i(k_1 x -\omega_1 t)}\,,~~
v(x,t) = B \sech(\xi) e^{i(k_2 x -\omega_2 t)}\,,
\ee
provided
\bea\label{1.2g}
&&\beta^2 = g_{11} A^2 + g_{12} B^2 = g_{21} A^2 + g_{22} B^2\,,~~
\omega_1+k_{1}^3 = -6 k_1 g_{11} A^2\,, 
\nonumber \\
&&c+3k_{1}^{2} = 4\beta^2 -6 g_{11} A^2\,,~~
\omega_2+k_{2}^3 = 3 k_2 \beta^2 -6 k_2 g_{21} A^2\,, 
\nonumber \\
&&c+3k_{2}^{2} = \beta^2 -6 g_{21} A^2\,.
\eea

{\bf Solution VII}

It is easy to check that
\be\label{1.2h}
u(x,t) = A \dn(\xi,m) e^{i(k_1 x -\omega_1 t)}\,,~~
v(x,t) = B \sqrt{m} \cn(\xi,m) e^{i(k_2 x -\omega_2 t)}\,,
\ee
is an exact solution of the coupled Eqs. (\ref{1.1}) and (\ref{1.2}) 
provided
\bea\label{1.2i}
&&\beta^2 = g_{11} A^2 + g_{12} B^2 = g_{21} A^2 + g_{22} B^2\,,
\nonumber \\
&&c+3k_{1}^{2} = (2-m)\beta^2 -6(1-m) g_{12} B^2\,,
\nonumber \\
&&\omega_2+k_{2}^3 = 3 k_2 \beta^2 -6(1-m) k_2 g_{22} B^2\,, 
\nonumber \\
&&c+3k_{2}^{2} = (5-4m)\beta^2 -6(1-m) g_{22} B^2\,,
\nonumber \\
&&\omega_1+k_{1}^3 = 3(2-m) k_1 \beta^2 -6(1-m) k_1 g_{12} A^2\,.
\eea

{\bf Solution VIII}

In the limit $m = 1$, the solution IV goes over to the hyperbolic 
solution
\be\label{1.2j}
u(x,t) = A \sech(\xi) e^{i(k_1 x -\omega_1 t)}\,,~~
v(x,t) = B \sech(\xi) e^{i(k_2 x -\omega_2 t)}\,,
\ee
provided
\bea\label{1.2k}
&&\beta^2 = g_{11} A^2 + g_{12} B^2 = g_{21} A^2 + g_{22} B^2\,,~~
\omega_1+k_{1}^3 = 3 k_1 \beta^2\,, 
\nonumber \\
&&c+3k_{1}^{2} = \beta^2\,,~~k_2 = \pm k_1\,,~~\omega_2 
= \pm \omega_1\,.
\eea
Note that the $\pm$ signs in Eq. (\ref{1.2k}) are correlated.

{\bf Solution IX}

It is straightforward to check that 
\be\label{1.2l}
u(x,t) = A \tanh(\xi) e^{i(k_1 x -\omega_1 t)}\,,~~
v(x,t) = B \tanh(\xi) e^{i(k_2 x -\omega_2 t)}\,,
\ee
is an exact solution of the coupled Eqs. (\ref{1.1}) and (\ref{1.2})
provided
\bea\label{1.2m}
&&\beta^2 = g_{11} A^2 + g_{12} B^2 = g_{21} A^2 + g_{22} B^2\,,~~
\omega_1+k_{1}^3 = -6 k_1 \beta^2\,, 
\nonumber \\
&&c+3k_{1}^{2} = -2\beta^2\,,~~k_2 = \pm k_1\,,~~\omega_2 
= \pm \omega_1\,.
\eea
Note that the $\pm$ signs in Eq. (\ref{1.2m}) are correlated. As mentioned
above, the corresponding local coupled mKdV Eqs. (\ref{1.5}) and 
(\ref{1.6}) do not admit any of the above six solutions unless, of course, 
if $k_1 = k_2 = \omega_1 = \omega_2 = 0$.

It easily follows from here that corresponding to any solution
of the uncoupled nonlocal Eq. (\ref{1.4g}) of the form 
\be\label{1.5f}
u(x,t) = u(\xi)\,,~~\xi = \beta(x-ct)\,,
\ee
where $u(\xi)$ is real, there is always a solution of the form 
\be\label{1.5g}
u(x,t) = u(\xi) e^{i(k_1 x-\omega_1 t)}\,,
\ee

We now show that the coupled Eqs. (\ref{1.1}) and (\ref{1.2}) admit 
 several new solutions which we present one by one. For simplicity,
 we present these solutions without the plane wave factor. However, we
 emphasize that all the coupled solutions discussed below are also valid
 with the plane wave factor as discussed in the above six solutions.

{\bf Solution X}

It is easy to check that 
\be\label{1.3}
u = \frac{A[F+\sech(\xi)]}{[D+\sech(\xi)]}\,,~~
v = \frac{B\sech(\xi)}{[D+\sech(\xi)]}\,,~~D > 0\,,
\ee
where $\xi = \beta(x-ct)$ is an exact solution of the coupled Eqs. 
(\ref{1.1}) and (\ref{1.2}) provided
\bea\label{1.4}
&&c(D-F) = (D+2F) \beta^2\,,~~g_{11} = g_{21}\,,~~g_{12} = g_{22}\,,
\nonumber \\
&&2g_{11} A^2 F(D-F) = D^2 \beta^2\,,~~2 g_{12} B^2 F 
= (2D^2 F-D-F)\beta^2\,.
\eea

It is worth pointing out that the corresponding local coupled mKdV 
Eqs. (\ref{1.5}) and (\ref{1.6})
also admit the solution (\ref{1.3}) provided the relations (\ref{1.4})
are satisfied.

{\bf Solution XI}

It is easy to check that 
\be\label{1.7}
u = \frac{A[F+\sech(\xi)]}{[D+\sech(\xi)]}\,,~~
v = \frac{B}{[D+\sech(\xi)]}\,,~~D > 0\,,
\ee
where $\xi = \beta(x-ct)$ is an exact solution of the coupled Eqs. 
(\ref{1.1}) and (\ref{1.2}) provided
\bea\label{1.8}
&&c(D-F) = (6FD^2-2D-F) \beta^2\,,~~g_{11} = g_{21}\,,~~g_{12} = g_{22}\,,
\nonumber \\
&&2g_{11} A^2 (D-F) = 3D(1-2D^2) \beta^2\,,~~2 g_{12} B^2 
= D(2D^2 F-D-F)\beta^2\,. \nonumber\\ 
\eea

It is worth pointing out that the corresponding local coupled mKdV 
Eqs. (\ref{1.5}) and (\ref{1.6}) also admit the solution (\ref{1.7}) 
provided the relations (\ref{1.8}) hold good.

{\bf Solution XII}

It is not difficult to check that 
\be\label{1.9}
u = \frac{A\sech(\xi)}{[D+\sech(\xi)]}\,,~~
v = \frac{B}{[D+\sech(\xi)]}\,,~~D > 0\,,
\ee
where $\xi = \beta(x-ct)$ is an exact solution of the coupled Eqs. 
(\ref{1.1}) and (\ref{1.2}) provided
\bea\label{1.10}
&&c = -2\beta^2\,,~~g_{11} = g_{21}\,,~~g_{12} = g_{22}\,,
\nonumber \\
&&2g_{11} A^2 = 3(2D^2-1) \beta^2\,,~~2 g_{12} B^2 
= -D^2 \beta^2\,.
\eea

It is worth pointing out that the corresponding local coupled mKdV 
Eqs. (\ref{1.5}) and (\ref{1.6}) also admit the solution (\ref{1.9}) 
provided the relations (\ref{1.10}) hold good.

{\bf Solution XIII}

It is easy to check that 
\be\label{1.11}
u = \frac{A[F+\cos(\xi)]}{[D+\cos(\xi)]}\,,~~
v = \frac{B\cos(\xi)}{[D+\cos(\xi)]}\,,~~D > 1\,,
\ee
where $\xi = \beta(x-ct)$ is an exact solution of the coupled Eqs. 
(\ref{1.1}) and (\ref{1.2}) provided
\bea\label{1.12}
&&D(F-D)c = (D^2-6+2DF) \beta^2\,,~~g_{11} = g_{21}\,,~~g_{12} = g_{22}\,,
\nonumber \\
&&2g_{11} A^2 F(F-D) = (D^2-2) \beta^2\,,~~2 g_{12} B^2 DF 
= (D^2+DF-2)\beta^2\,. \nonumber\\ 
\eea

It is worth pointing out that the corresponding local coupled mKdV 
Eqs. (\ref{1.5}) and (\ref{1.6}) also admit the solution (\ref{1.11}) 
provided the relations (\ref{1.12}) hold good.

{\bf Solution XIV}

It is straightforward to check that 
\be\label{1.13}
u = \frac{A[F+\cos(\xi)]}{[D+\cos(\xi)]}\,,~~
v = \frac{B}{[D+\cos(\xi)]}\,,~~D > 1\,,
\ee
where $\xi = \beta(x-ct)$ is an exact solution of the coupled Eqs. 
(\ref{1.1}) and (\ref{1.2}) provided
\bea\label{1.14}
&&(D-F)c = (2D+F) \beta^2\,,~~g_{11} = g_{21}\,,~~g_{12} = g_{22}\,,
\nonumber \\
&&2g_{11} A^2 (D-F) = D \beta^2\,,~~2 g_{12} B^2 
= (D^2+DF-2)\beta^2\,.
\eea

It is worth pointing out that the corresponding local coupled mKdV 
Eqs. (\ref{1.5}) and (\ref{1.6}) also admit the solution (\ref{1.13}) 
provided the relations (\ref{1.14}) hold good.

{\bf Solution XV}

It is easy to check that
\be\label{1.15}
u = \frac{A\cos(\xi)}{[D+\cos(\xi)]}\,,~~
v = \frac{B}{[D+\cos(\xi)]}\,,~~D > 1\,,
\ee
where $\xi = \beta(x-ct)$ is an exact solution of the coupled Eqs. 
(\ref{1.1}) and (\ref{1.2}) provided
\bea\label{1.16}
&&c = 2 \beta^2\,,~~g_{11} = g_{21}\,,~~g_{12} = g_{22}\,,
\nonumber \\
&&2g_{11} A^2  =  \beta^2\,,~~2 g_{12} B^2 
= (D^2-2)\beta^2\,.
\eea

It is worth pointing out that the corresponding local coupled mKdV 
Eqs. (\ref{1.5}) and (\ref{1.6}) also admit the solution (\ref{1.15}) 
provided the relations (\ref{1.16}) hold good.

{\bf Solution XVI}

It is not difficult to check that 
\be\label{1.17}
u = \frac{A[F+\tanh(\xi)]}{[D+\tanh(\xi)]}\,,~~
v = \frac{B\tanh(\xi)}{[D+\tanh(\xi)]}\,,~~D > 1\,,
\ee
where $\xi = \beta(x-ct)$ is an exact solution of the coupled Eqs. 
(\ref{1.1}) and (\ref{1.2}) provided
\bea\label{1.18}
&&cD(F-D) = 2(2DF+D^2-3) \beta^2\,,~~g_{11} = g_{21}\,,~~g_{12} = g_{22}\,,
\nonumber \\
&&g_{11} A^2 F(F-D) = (D^2-1) \beta^2\,,~~ g_{12} B^2 DF 
= (D F +D^2-1-D^3 F)\beta^2\,. \nonumber\\ 
\eea

It is worth pointing out that the corresponding local coupled mKdV 
Eqs. (\ref{1.5}) and (\ref{1.6}) also admit the solution (\ref{1.17}) 
provided the relations (\ref{1.18}) hold good.

{\bf Solution XVII}

It is easy to check that 
\be\label{1.19}
u = \frac{A[F+\tanh(\xi)]}{[D+\tanh(\xi)]}\,,~~
v = \frac{B}{[D+\tanh(\xi)]}\,,~~D > 1\,,
\ee
where $\xi = \beta(x-ct)$ is an exact solution of the coupled Eqs. 
(\ref{1.1}) and (\ref{1.2}) provided
\bea\label{1.20}
&&c(D-F) = 2(2D+F-3F D^2) \beta^2\,,~~g_{11} = g_{21}\,,~~g_{12} = g_{22}\,,
\nonumber \\
&&g_{11} A^2 (D-F) = D(1-D^2) \beta^2\,,~~ g_{12} B^2 
= (D^2-1)(1-DF)\beta^2\,.~~~
\eea

It is worth pointing out that the corresponding local coupled mKdV 
Eqs. (\ref{1.5}) and (\ref{1.6}) also admit the solution (\ref{1.19}) 
provided the relations (\ref{1.20}) hold good.

{\bf Solution XVIII}

It is straightforward to check that 
\be\label{1.21}
u = \frac{A[F+\dn(\xi,m)]}{[D+\dn(\xi,m)]}\,,~~
v = \frac{B\dn(\xi,m)}{[D+\dn(\xi,m)]}\,,~~D > 0\,,
\ee
where $\xi = \beta(x-ct)$ is an exact solution of the coupled Eqs. 
(\ref{1.1}) and (\ref{1.2}) provided
\bea\label{1.22}
&&cD(D-F) = [(2-m)D(D+2F)-6(1-m)] \beta^2\,,~~g_{11} = g_{21}\,,
~~g_{12} = g_{22}\,,
\nonumber \\
&&2g_{11} A^2 F(D-F) = [(2-m)D^2-2(1-m)] \beta^2\,, \nonumber \\
&&2 g_{12} B^2 DF = [2(1-m)-(2-m)D(D+F)+2D^3F]\beta^2\,.
\eea

It is worth pointing out that the corresponding local coupled mKdV 
Eqs. (\ref{1.5}) and (\ref{1.6}) also admit the solution (\ref{1.21}) 
provided the relations (\ref{1.22}) hold good.

{\bf Solution XIX}

It is easy to check that 
\be\label{1.23}
u = \frac{A[F+\dn(\xi,m)]}{[D+\dn(\xi,m)]}\,,~~
v = \frac{B}{[D+\dn(\xi,m)]}\,,~~D > 0\,,
\ee
where $\xi = \beta(x-ct)$ is an exact solution of the coupled Eqs. 
(\ref{1.1}) and (\ref{1.2}) provided
\bea\label{1.24}
&&c(D-F) = [6FD^2-(2-m)(2D+F)] \beta^2\,,~~g_{11} = g_{21}\,,~~g_{12} 
= g_{22}\,,
\nonumber \\
&&2g_{11} A^2 (D-F) = D[2D^2-(2-m)] \beta^2\,, \nonumber \\
&&2 g_{12} B^2 = [2(1-m)-(2-m)D(D+F)-2D^3 F]\beta^2\,.
\eea

It is worth pointing out that the corresponding local coupled mKdV 
Eqs. (\ref{1.5}) and (\ref{1.6}) also admit the solution (\ref{1.23}) 
provided the relations (\ref{1.24}) hold good.

{\bf Solution XX}

It is not difficult to check that 
\be\label{1.25}
u = \frac{A\dn(\xi,m)}{[D+\dn(\xi,m)]}\,,~~
v = \frac{B}{[D+\dn(\xi,m)]}\,,~~D > 0\,,
\ee
where $\xi = \beta(x-ct)$ is an exact solution of the coupled Eqs. 
(\ref{1.1}) and (\ref{1.2}) provided
\bea\label{1.26}
&&c = -2(2-m)\beta^2\,,~~g_{11} = g_{21}\,,~~g_{12} = g_{22}\,,
\nonumber \\
&&2g_{11} A^2 = [2D^2-(2-m)] \beta^2\,,~~2 g_{12} B^2 
= [2(1-m)-(2-m)D^2] \beta^2\,. \nonumber\\ 
\eea

It is worth pointing out that the corresponding local coupled mKdV 
Eqs. (\ref{1.5}) and (\ref{1.6}) also admit the solution (\ref{1.25}) 
provided the relations (\ref{1.26}) hold good.

{\bf Solution XXI}

It is easy to check that 
\be\label{1.27}
u = \frac{A[F+\cn(\xi,m)]}{[D+\cn(\xi,m)]}\,,~~
v = \frac{B\cn(\xi,m)}{[D+\cn(\xi,m)]}\,,~~D > 1\,,
\ee
where $\xi = \beta(x-ct)$ is an exact solution of the coupled Eqs. 
(\ref{1.1}) and (\ref{1.2}) provided
\bea\label{1.28}
&&D(D-F)c = [6(1-m)+(2m-1)D(D+2F)] \beta^2\,,~~g_{11} = g_{21}\,,
~~g_{12} = g_{22}\,,
\nonumber \\
&&2g_{11} A^2 F(D-F) = [2(1-m)+(2m-1)D^2] \beta^2\,, \nonumber \\
&&2 g_{12} B^2 DF = [2mD^3F-(2m-1)D(D+F)-2(1-m)]\beta^2\,. 
\eea

It is worth pointing out that the corresponding local coupled mKdV 
Eqs. (\ref{1.5}) and (\ref{1.6}) also admit the solution (\ref{1.27}) 
provided the relations (\ref{1.28}) hold good.

{\bf Solution XXII}

It is easy to check that 
\be\label{1.29}
u = \frac{A[F+\cn(\xi,m)]}{[D+\cn(\xi,m)]}\,,~~
v = \frac{B}{[D+\cn(\xi,m)]}\,,~~D > 1\,,
\ee
where $\xi = \beta(x-ct)$ is an exact solution of the coupled Eqs. 
(\ref{1.1}) and (\ref{1.2}) provided
\bea\label{1.30}
&&(D-F)c = [6mF D^2-(2m-1)(2D+F)] \beta^2\,,~~g_{11} = g_{21}\,,
~~g_{12} = g_{22}\,,
\nonumber \\
&&2g_{11} A^2 (D-F) = D[2m-1-2mD^2] \beta^2\,, \nonumber \\
&&2 g_{12} B^2 = -[2(1-m)+D(F+D)+2mD^3F]\beta^2\,.
\eea

It is worth pointing out that the corresponding local coupled mKdV 
Eqs. (\ref{1.5}) and (\ref{1.6}) also admit the solution (\ref{1.29}) 
provided the relations (\ref{1.30}) hold good.

{\bf Solution XXIII}

It is easy to check that 
\be\label{1.31}
u = \frac{A\cos(\xi)}{[D+\cos(\xi)]}\,,~~
v = \frac{B}{[D+\cos(\xi)]}\,,~~D > 1\,,
\ee
where $\xi = \beta(x-ct)$ is an exact solution of the coupled Eqs. 
(\ref{1.1}) and (\ref{1.2}) provided
\bea\label{1.32}
&&c = -2(2m-1) \beta^2\,,~~g_{11} = g_{21}\,,~~g_{12} = g_{22}\,,
\nonumber \\
&&2g_{11} A^2  =  [2mD^2-(2m-1)]\beta^2\,,~~2 g_{12} B^2 
= -[(2m-1)D^2+2(1-m)]\beta^2\,. \nonumber\\ 
\eea

It is worth pointing out that the corresponding local coupled mKdV 
Eqs. (\ref{1.5}) and (\ref{1.6}) also admit the solution (\ref{1.31}) 
provided the relations (\ref{1.32}) hold good.

We now present six soliton solutions with a power law tail.

{\bf Solution XXIV}

It is not difficult to check that 
\be\label{1.33}
u = \frac{A(F+x^2)}{D+x^2}\,,~~
v = \frac{B x^2}{D+ x^2}\,,~~D > 0\,,
\ee
where $\xi = \beta(x-ct)$ is an exact solution of the coupled Eqs. 
(\ref{1.1}) and (\ref{1.2}) provided
\bea\label{1.34}
&&cD(F-D) = 6(F+2D)\,,~~g_{11} = g_{21}\,,~~g_{12} = g_{22}\,,
\nonumber \\
&& F(F-D) g_{11} A^2  = 3D\,,~~ F D g_{12} B^2 = F+3D\,.
\eea

It is worth pointing out that the corresponding local coupled mKdV 
Eqs. (\ref{1.5}) and (\ref{1.6}) also admit the solution (\ref{1.33}) 
provided the relations (\ref{1.34}) hold good.

{\bf Solution XXV}

It is straightforward to check that 
\be\label{1.35}
u = \frac{A(F+x^2)}{D+x^2}\,,~~
v = \frac{B}{D+ x^2}\,,~~D > 0\,,
\ee
where $\xi = \beta(x-ct)$ is an exact solution of the coupled Eqs. 
(\ref{1.1}) and (\ref{1.2}) provided
\bea\label{1.36}
&&c(D-F) = 6\,,~~g_{11} = g_{21}\,,~~g_{12} = g_{22}\,,
\nonumber \\
&& (F-D) g_{11} A^2  = 1\,,~~(D-F) g_{12} B^2 = F+3D\,.
\eea

It is worth pointing out that the corresponding local coupled mKdV 
Eqs. (\ref{1.5}) and (\ref{1.6}) also admit the solution (\ref{1.35}) 
provided the relations (\ref{1.36}) hold good.

{\bf Solution XXVI}

It is easy to check that 
\be\label{1.37}
u = \frac{A x^2}{D+x^2}\,,~~
v = \frac{B}{D+ x^2}\,,~~D > 0\,,
\ee
where $\xi = \beta(x-ct)$ is an exact solution of the coupled Eqs. 
(\ref{1.1}) and (\ref{1.2}) provided
\bea\label{1.38}
&&c D = 6\,,~~g_{11} = g_{21}\,,~~g_{12} = g_{22}\,,
\nonumber \\
&& D g_{11} A^2  = 1\,,~~ g_{12} B^2 = 3D\,.
\eea

It is worth pointing out that the corresponding local coupled mKdV 
Eqs. (\ref{1.5}) and (\ref{1.6}) also admit the solution (\ref{1.37}) 
provided the relations (\ref{1.38}) hold good.

{\bf Solution XXVII}

It is not difficult to check that 
\be\label{6.45}
u = \frac{A(F+x^2)}{D+x^2}\,,~~
v = \frac{B x}{\sqrt{D+ x^2}}\,,~~D > 0\,,
\ee
where $\xi = \beta(x-ct)$ is an exact solution of the coupled Eqs. 
(\ref{1.1}) and (\ref{1.2}) provided
\bea\label{6.46}
&&cD(F-D)^2 = 12(F^2+2DF-D^2) = 3D(4F^2-D^2+2FD)\,, \nonumber \\
&&(F-D)^2 g_{11} A^2 = 4D\,,~~D(F-D) g_{12} B^2 = 2(F+3D)\,,  \nonumber \\
&& (F-D)^2 g_{21} A^2  = (5/2) D^2\,,~~ (F-D) g_{22} B^2 = 2F+3D\,.
\eea
On equating the two expressions for $c$ we find that $D \ne 4$ and further
if $D=1$ then $F =1/2$. Moreover, if $D \ne 1, 4$, then $F$ is given in 
terms of $D$ by
\be\label{6.47}
F = \frac{D(4-D) \pm D\sqrt{32-28D+5D^2}}{4(D-1)}\,.
\ee
Notice that $F$ is real provided $D < 8/5$ or $D > 4$.

{\bf Solution XXVIII}

It is easy to check that 
\be\label{6.48}
u = \frac{A x^2}{D+x^2}\,,~~
v = \frac{Bx}{\sqrt{D+ x^2}}\,,~~D > 0\,,
\ee
where $\xi = \beta(x-ct)$ is an exact solution of the coupled Eqs. 
(\ref{1.1}) and (\ref{1.2}) provided
\bea\label{3.49}
&&c = -3\,,~~D = 4\,,~~g_{11} A^2 =1\,,~~g_{12} B^2 = -3/2\,,
\nonumber \\
&& g_{21} A^2  = 5/2\,,~~g_{22} B^2 = -3\,.
\eea

{\bf Solution XXIX}

It is straightforward to check that 
\be\label{6.50}
u = \frac{A}{D+x^2}\,,~~
v = \frac{B}{\sqrt{D+ x^2}}\,,~~D > 0\,,
\ee
where $\xi = \beta(x-ct)$ is an exact solution of the coupled Eqs. 
(\ref{1.1}) and (\ref{1.2}) provided
\bea\label{6.51}
&&c = 12\,,~~D = 1\,,~~g_{11} A^2 = 4\,,~~g_{12} B^2 = 2\,,
\nonumber \\
&& g_{21} A^2  = 5/2\,,~~ g_{22} B^2 = 2\,.
\eea

\section{Conclusion And Open Problems}

In this paper we have obtained several new solutions of the coupled AM variant 
of the nonlocal NLS equation as well as the coupled mKdV equations.  Moreover, 
we compared and contrasted these solutions with the solutions (in case they are 
admitted) of the corresponding coupled NLS or mKdV equations, respectively. In  
the nonlocal NLS case, particular mention may be made of the host of the periodic 
as well as the hyperbolic nonreciprocal solutions in a special subclass of the coupled 
nonlocal NLS solutions. In the nonlocal mKdV case, we have shown that in 
contrast to the local mKdV case, the nonlocal coupled mKdV equations 
admit plane wave solutions as well as moving soliton solutions. This paper
raises several open questions, some of which are as follows. 

\begin{enumerate}

\item While we have obtained a large number of nonreciprocal solutions
for the coupled nonlocal (as well as local) NLS equations, we have so far
not been able to obtain similar solutions for either the coupled nonlocal or
the local mKdV equations. It is worthwhile asking if nonreciprocal 
solutions can also be obtained for the mKdV or other coupled nonlocal 
or even local nonlinear equations.

\item Unlike the local mKdV equations we have shown that the nonlocal mKdV 
equations admit plane wave solutions as well as soliton solutions multiplied
by the plane wave factor. It is worth
asking if similar solutions can also be obtained for the other nonlocal 
equations. 

\item For the full coupled nonlocal NLS equations as well as coupled 
nonlocal mKdV equations we have obtained novel periodic as well as 
hyperbolic solutions. One obvious question is, can one similarly obtain 
exact solutions of the other coupled nonlocal equations, e.g., coupled 
Hirota or coupled KdV equations or coupled Yang-type nonlocal NLS? 
Further, in the coupled nonlocal KdV case do we again get plane wave 
solutions as well as soliton solutions multiplied by the plane wave factor? 

\end{enumerate}

\section{Acknowledgment} 

One of us (AK) is grateful to Indian National Science Academy (INSA) for the 
award of INSA Honorary Scientist position at Savitribai Phule Pune University. 
The work at Los Alamos National Laboratory was carried out under the auspices 
of the US DOE and NNSA under contract No.~DEAC52-06NA25396.

\section{\bf Appendix A: Novel Periodic Superposed Solutions of a Coupled 
Local NLS Model}

We now present those solutions of the coupled local NLS Eqs. (\ref{6})
and (\ref{7}) which are, however, not the solutions of the nonlocal coupled
Eqs. (\ref{1}) and (\ref{2}).

In this context, it is interesting to observe that unlike the coupled
nonlocal AM Eqs. (\ref{1}) and (\ref{2}), the (local) coupled NLS 
Eqs. (\ref{6}) and (\ref{7}) admit the plane wave solution
\be\label{A50}
u(x,t) = A e^{i(\omega_1 t-k_1 x)}\,,~~v(x,t) 
= B e^{i(\omega_2 t-k_2)}\,,
\ee
provided
\bea\label{A51}
&&\omega_1 + k_{1}^2 = [g_{11} A^2 + g_{12} B^2]\,,
\nonumber \\
&&\omega_2 + k_{2}^2 = [g_{21} A^2 + g_{22} B^2]\,.
\eea

We now show that corresponding to any stationary soliton solution of the
coupled local NLS Eqs. (\ref{6}) and (\ref{7}), one always has a 
corresponding moving soliton solution. The proof is straightforward. 
Let us assume that there is a stationary soliton solution of the form
\be\label{A52}
u(x,t) = u(\beta x) e^{i(\omega_1 t-k_1 x}\,,~~v(x,t) = v(\beta x) 
e^{i(\omega_2 t -k_2 x)}\,,~~\xi = \beta(x-ct)\,,
\ee
which is an exact solution of the coupled Eqs. (\ref{6}) and (\ref{7}). Using
the ansatz (\ref{A52}) in Eqs. (\ref{6}) and (\ref{7}), we find that it 
is an exact solution provided
\bea\label{A53}
&&\beta^2 u''(\eta) = \omega_1 u
+[g_{11}|u|^2+g_{12}|v|^2]u\,, \nonumber \\
&&\beta^2 v''(\eta) = \omega_2 v
+[g_{21}|u|^2+g_{22}|v|^2]v\,.
\eea

Let us now consider the moving soliton solutions. Let 
\be\label{A54}
u(x,t) = u(\xi) e^{i(\omega_1 t-k_1 x)}\,,~~v(x,t) = v(\xi) 
e^{i(\omega_2 t -k_2 x)}\,,~~\xi = \beta(x-ct)\,,
\ee
be an exact solution of the coupled local NLS Eqs. (\ref{6}) and (\ref{7}).
On substituting this ansatz in Eqs. (\ref{6}) and (\ref{7}), we find that 
it is an exact solution provided
\bea\label{A55}
&&k_1 = k_2\,,~~c = -2k_1\,,~~\beta^2 u''(\xi) = (\omega_1+k_{1}^2) u
+[g_{11}|u|^2+g_{12}|v|^2]u\,, \nonumber \\
&&\beta^2 v''(\xi) = (\omega_2+k_{1}^2) v
+[g_{21}|u|^2+g_{22}|v|^2]v\,.
\eea
Now notice that in the limit $k_1 = k_2 = c = 0$, the coupled Eqs. (\ref{A55}) 
indeed go over to Eqs. (\ref{A53}). 

For simplicity, we now only present 
the new stationary soliton solutions of the local coupled NLS Eqs. (\ref{6}) 
and (\ref{7}). But as proved above, it is understood that there are 
corresponding moving soliton solutions.

{\bf Solution IA}

It is not difficult to check that
\be\label{A1}
u(x,t) = e^{i\omega_1 t} \frac{A\dn(\beta x,m)}{D+\sn(\beta x, m)}\,,~~
v(x,t) = e^{i\omega_2 t} \frac{B\cn(\beta x,m)}{D+\sn(\beta x, m)}\,,
\ee
with $A, B > 0\,, D > 1$, is an exact solution of the coupled 
Eqs. (\ref{6}) and (\ref{7}) provided
\bea\label{A2}
&&\omega_1 = \omega_2 = \frac{(1+m)\beta^2}{2}\,,~~g_{12} B^2 = 
\frac{3(mD^2-1)\beta^2}{2}\,,~~g_{11} A^2 = \frac{(D^2-1)\beta^2}{2}\,, 
\nonumber \\
&&g_{22} B^2 = \frac{(mD^2-1)\beta^2}{2}\,,~~g_{21} A^2 = 
\frac{3(D^2-1)\beta^2}{2}\,.
\eea
Thus $g_{12}, g_{22} > 0$ if $1 < D^2 < 1/m$ 
while $g_{11}, g_{21} > 0$. On the other hand $\omega_1, \omega_2 > 0$.

{\bf Solution IIA}

It is easy to check that 
\be\label{A3}
u(x,t) = e^{i\omega_1 t} \frac{A[F+\tanh(\beta x)]}{D+\tanh(\beta x)}\,,~~
v(x,t) = e^{i\omega_2 t} \frac{B\sech(\beta x)}{D+\tanh(\beta x)}\,,
\ee
with $A, B > 0, D > 1$, is an exact solution of the coupled 
local NLS Eqs. (\ref{6}) and (\ref{7}) (but not nonlocal coupled Eqs.
(\ref{1}) and (\ref{2})) provided
\bea\label{A4}
&&\omega_1 = 4\beta^2\,,~~\omega_2 = \beta^2\,,~~F = \pm 1\,,~~g_{11} A^2 
= (D \pm 1)^2 \beta^2\,, \nonumber \\
&&g_{12} B^2 = 2(D^2-1)\beta^2\,,~~g_{22} = g_{12}\,,~~g_{21} = 0\,.
\eea
Note that the $\pm$ signs here are correlated.   

{\bf Solution IIIA}

It is straightforward to check that
\be\label{A5}
\phi_1 = A\left[\frac{1-m}{\dn^2(\beta x,m)}+y\right]
+\frac{B\sqrt{m}(1-m)\sn(\beta x, m)}{\dn^2(\beta x, m)}\,,
\ee
\be\label{A6}
\phi_2 = \frac{D\sqrt1-{m}}{\dn(\beta x,m)}
+\frac{E\sqrt{m(1-m)}\sn(\beta x, m)}{\dn(\beta x, m)}\,,
\ee
is an exact nonreciprocal solution of the unusual coupled local NLS Eqs. 
(\ref{8.1a}) and (\ref{8.2a}) (but not the nonlocal Eqs. (\ref{8.1}) and
(\ref{8.2})) provided $0 < m < 1$ and further
\bea\label{A7}
&&d_1 D^2 = -\frac{3\beta^2}{2}\,,~~a_1 = [\frac{(7+m)}{2}+3(1-m)y]\beta^2\,,~~
E = \pm D\,,~~B = \pm A\,, \nonumber \\
&&a_2 = \frac{(1+m)\beta^2}{2}\,,~~ d_2 D^2 = -\frac{\beta^2}{2}\,,~~
y = \frac{-(5-m) \pm \sqrt{1+14m+m^2}}{6}\,.
\eea
Note that the $\pm$ signs in $D, E$ are correlated with the signs in 
$B$ and $A$. Further, the solution is valid for arbitrary values of $A$.

{\bf Solution IVA}

It is easy to check that
\be\label{A8}
\phi_1 = \frac{A\sqrt{m}\cn(\beta x, m)}{\dn^2(\beta x,m)} 
+\frac{B m \cn(\beta x, m) \sn(\beta x, m)}{\dn^2(\beta x, m)}\,, 
\ee
\be\label{A9}
\phi_2 = \frac{D\sqrt{1-m}}{\dn(\beta x,m)}
+\frac{E\sqrt{m(1-m)}\sn(\beta x, m)}{\dn(\beta x, m)}\,,
\ee
is an exact nonreciprocal solution of the unusual coupled local NLS Eqs. 
(\ref{8.1a}) and (\ref{8.2a}) (but not the solutions of the coupled nonlocal
Eqs. (\ref{8.1}) and (\ref{8.2})) provided
\bea\label{A10}
&&d_1 D^2 = -\frac{3\beta^2}{2}\,,~~a_1 = a_2 = \frac{(1+m)\beta^2}{2}\,,~~
E = \pm D\,,~~B = \pm A\,, \nonumber \\
&&d_2 D^2 = -\frac{\beta^2}{2}\,,~~0 < m < 1\,.
\eea
Note that the $\pm$ signs in $D, E$ are correlated with the signs in 
$B$ and $A$. Further, the solution is valid for arbitrary values of $A$.

We now present three solutions of the coupled local NLS where in general 
$k_1 \ne k_2$. Further, in these cases even solutions with $c = 0$ are 
allowed even though $k_1, k_2$ are still nonzero and in general unequal.

{\bf Solution VA}

It is not difficult to check that 
\bea\label{A28}
&&u(x,t) = \sqrt{n_1}[\cos(\theta_1)\tanh(\xi)+i\sin(\theta_1)] 
e^{i(k_1 x- \omega_1 t)}\,, \nonumber \\
&&v(x,t) = \sqrt{n_2}[\cos(\theta_2)\tanh(\xi)+i\sin(\theta_2)] 
e^{i(k_2 x- \omega_2 t)}\,,~~\xi = \beta(x-ct)\,, \nonumber\\ 
\eea
is an exact solution of the local NLS coupled Eqs. (\ref{6}) and (\ref{7})
provided
\be\label{A29}
n_1 g_{11} \cos^2(\theta_1) +n_2 g_{12} \cos^2(\theta_2)  
= n_1 g_{21} \cos^2(\theta_1) + n_2 g_{22} \cos^2(\theta_2) = -2\beta^2\,,
\ee
\be\label{A30}
\omega_1 -k_{1}^2 + n_1 g_{11} + n_2 g_{12} = 0\,,
\ee
\be\label{A31}
\omega_2 -k_{2}^2 + n_1 g_{21} + n_2 g_{22} = 0\,,
\ee
\be\label{A32}
c = 2k_1  +2\beta \tan(\theta_1) = 2k_2 +2\beta \tan(\theta_2)\,. 
\ee

{\bf Solution VIA}

It is easy to check that 
\bea\label{A33}
&&u(x,t) = \sqrt{n_1}[\cos(\theta_1)\tanh(\xi)+i\sin(\theta_1)] 
e^{i(k_1 x- \omega_1 t)}\,, \nonumber \\
&&v(x,t) = \sqrt{n_2}[\sin(\theta_2)+i\cos(\theta_2)\tanh(\xi)] 
e^{i(k_2 x- \omega_2 t)}\,,~~\xi = \beta(x-ct)\,, \nonumber\\ 
\eea
is an exact solution of the local NLS coupled Eqs. (\ref{6}) and (\ref{7})
provided Eqs. (\ref{A29}) to Eqs. (\ref{A31}) are satisfied and further
\be\label{A34}
c = 2k_1  +2\beta \tan(\theta_1) = 2k_2 - 2\beta \tan(\theta_2)\,. 
\ee

{\bf Solution VIIA}

It is straightforward to check that 
\bea\label{A35}
&&u(x,t) = \sqrt{n_1}[\sin(\theta_1)+i\cos(\theta_1)\tanh(\xi)] 
e^{i(k_1 x- \omega_1 t)}\,, \nonumber \\
&&v(x,t) = \sqrt{n_2}[\sin(\theta_2)+i\cos(\theta_2)\tanh(\xi)] 
e^{i(k_2 x- \omega_2 t)}\,,~~\xi = \beta(x-ct)\,, \nonumber\\ 
\eea
is an exact solution of the local NLS coupled Eqs. (\ref{6}) and (\ref{7})
provided Eqs. (\ref{A29}) to Eqs. (\ref{A31}) are satisfied and further
\be\label{A36}
c = 2k_1  - 2\beta \tan(\theta_1) = 2k_2 - 2\beta \tan(\theta_2)\,. 
\ee

\section{\bf Appendix B: Novel Periodic Superposed Solutions of a Coupled 
Local mKdV Model}

We now present those solutions of the coupled local mKdV Eqs. (\ref{1.5})
and (\ref{1.6}) which are however not the solutions of the coupled 
nonlocal Eqs. (\ref{1.1}) and (\ref{1.2}).

{\bf Solution IB}

It is easy to check that 
\be\label{B1}
u(x,t) = \frac{A\tanh(\xi)}{[D+\tanh(\xi)]}\,,~~
v(x,t) = \frac{B}{[D+\tanh(\xi)]}\,,~~D > 1\,,
\ee
where $\xi = \beta(x-ct)$ is an exact solution of the coupled Eqs. 
(\ref{1.5}) and (\ref{1.6}) provided
\bea\label{B2}
&&c = 4\beta^2\,,~~g_{11} = g_{21}\,,~~g_{12} = g_{22}\,,
\nonumber \\
&&g_{11} A^2 = (1-D^2) \beta^2\,,~~ g_{12} B^2 
= (D^2-1) \beta^2\,.
\eea

{\bf Solution IIB}

It is not difficult to check that 
\be\label{B3}
u(x,t) = \frac{A[F+\sn(\xi,m)]}{[D+\sn(\xi,m)]}\,,~~
v(x,t) = \frac{B\sn(\xi,m)}{[D+\sn(\xi,m)]}\,,~~D > 1\,,
\ee
where $\xi = \beta(x-ct)$ is an exact solution of the coupled Eqs. 
(\ref{1.5}) and (\ref{1.6}) provided
\bea\label{B4}
&&cD(F-D) = 2[(1+m)D(D+2F)-3] \beta^2\,,~~g_{11} = g_{21}\,,~~g_{12} = g_{22}\,,
\nonumber \\
&&2g_{11} A^2 F(F-D) = [(1+m)D^2-2] \beta^2\,, \nonumber \\
&&2g_{12} B^2 DF = [(1+m)D(D+F)-2-2mD^3 F]\beta^2\,.  
\eea

{\bf Solution IIIB}

It is easy to check that 
\be\label{B5}
u(x,t) = \frac{A[F+\sn(\xi,m)]}{[D+\sn(\xi,m)]}\,,~~
v(x,t) = \frac{B}{[D+\sn(\xi,m)]}\,,~~A, B > 0\,,~~D > 1\,,
\ee
where $\xi = \beta(x-ct)$ is an exact solution of the coupled Eqs. 
(\ref{1.5}) and (\ref{1.6}) provided
\bea\label{B6}
&&c(D-F) = [(2D+F)(1+m)-6m F D^2] \beta^2\,,~~g_{11} = g_{21}\,,
~~g_{12} = g_{22}\,, \nonumber \\
&&2g_{11} A^2 (D-F) = D[1+m-2m D^2) \beta^2\,, \nonumber \\
&&2g_{12} B^2 = [(1+m)(D^2+DF)-2-2m F D^3]\beta^2\,. 
\eea

{\bf Solution IVB}

It is straightforward to check that 
\be\label{B7}
u(x,t) = \frac{A\sn(\xi,m)}{[D+\sn(\xi,m)]}\,,~~
v(x,t) = \frac{B}{[D+\sn(\xi,m)]}\,,~~D > 1\,,
\ee
where $\xi = \beta(x-ct)$ is an exact solution of the coupled Eqs. 
(\ref{1.5}) and (\ref{1.6}) provided
\bea\label{6.26}
&&c = 2(1+m)\beta^2\,,~~g_{11} = g_{21}\,,~~g_{12} = g_{22}\,,
\nonumber \\
&&2g_{11} A^2 = (1+m-2m D^2) \beta^2\,,~~ 2g_{12} B^2 
= [(1+m)D^2-2] \beta^2\,.
\eea


\end{document}